\documentclass[a4paper,11pt]{article}
\usepackage{jcappub} 
\pdfoutput=1 

\usepackage{natbib}
\usepackage{aas_macros}

\usepackage[T1]{fontenc} 
\usepackage{graphicx}
\usepackage{epsfig}
\usepackage{rotate}
\usepackage{amsmath}
\usepackage{amssymb}
\usepackage{amsfonts}
\usepackage{bm,outlines}
\usepackage{xspace}
\usepackage{diagbox}
\usepackage{slashbox}
\usepackage{verbatim}
\usepackage{tablefootnote}
\usepackage{enumerate}
\usepackage{afterpage}
\usepackage{xcolor}

\usepackage{hhline}

\usepackage{hyperref}

\usepackage[capitalise]{cleveref}
\Crefname{figure}{Fig.}{Figs.}
\Crefname{equation}{Eq.}{Eqs.}

\newcommand{\kpar}{k_{\parallel}}
\newcommand{\kparmin}{k_{\parallel,\, {\rm min}}}
\newcommand{\kperp}{k_{\bot}}

\newcommand{\bk}{\bm{k}}

\newcommand{\fnl}{f_{\rm NL}}

\newcommand*{\veps} {\varepsilon}
\newcommand*{\ie} {i.\,\!e.\ }

\makeatletter
\gdef\@fpheader{}
\makeatother

\title{Searching for primordial features with radio surveys: synergy between the power spectrum and bispectrum}

\author[1,2,3,5]{Dionysios Karagiannis,}
\author[1,2,3,4]{Mario Ballardini,}
\author[3,6]{Roy Maartens}

\affiliation[1]{Dipartimento di Fisica e Scienze della Terra, Universit\`a degli Studi di Ferrara, Via Giuseppe Saragat 1, 44122 Ferrara, Italy}
\affiliation[2]{INFN, Sezione di Ferrara, Via Giuseppe Saragat 1, 44122 Ferrara, Italy}
\affiliation[3]{Department of Physics \& Astronomy, University of the Western Cape, Cape Town 7535, South Africa}
\affiliation[4]{INAF-Osservatorio di Astrofisica e Scienza dello Spazio di Bologna, Via Piero Gobetti 93/3, 40129 Bologna, Italy}
\affiliation[5]{Van Swinderen Institute, University of Groningen, Nijenborgh 3, 9747 AG Groningen, The Netherlands}
\affiliation[6]{National Institute for Theoretical and Computational Sciences, Cape Town 7535, South Africa}
\emailAdd{d.karagiannis@rug.nl}
\emailAdd{mario.ballardini@unife.it}
\emailAdd{rmaartens@uwc.ac.za}

\abstract{\small We present a comprehensive forecasting framework to assess the detection of primordial oscillatory features by exploiting the synergy between future neutral hydrogen (HI) intensity mapping (IM) surveys and cosmic microwave background (CMB) measurements. Focusing on next-generation single-dish (SKAO) and interferometric (HIRAX) radio configurations, we perform a joint analysis of the redshift-space power spectrum and bispectrum, consistently incorporating scale-dependent bias, redshift-space distortions, and non-Gaussian covariance. We investigate phenomenological templates with linear and logarithmic primordial oscillations, together with a physically motivated sharp-feature model. We find that including the large-scale structure bispectrum improves marginalised constraints on feature amplitudes by $30$--$40\%$ relative to the power spectrum alone and helps break parameter degeneracies. In several cases, the bispectrum contains more information on the power-spectrum feature parameters than the power spectrum itself. Much of this gain arises from the late-time gravitational contribution, which inherits the oscillatory structure of the primordial feature signal and acts as an independent source of information. While the CMB angular power spectrum is crucial for constraining low oscillation frequencies, joint interferometric IM analyses ($P+B$) outperform CMB amplitude constraints by up to $75\%$ in the linear regime. We also show that, despite non-Gaussian covariance degrading the independent constraining power of the bispectrum by $55$--$90\%$, the combined HI+CMB probe achieves percent-level precision on the frequency of primordial features, providing a powerful test of non-slow-roll inflation.}

\graphicspath{{./plots/}}

\begin{document}

\maketitle

\section{Introduction} \label{sec:intro}
The standard model of cosmology, supported by the paradigm of cosmic inflation, has been remarkably successful in describing the statistical properties of the Universe on large scales. Current observations, most notably the measurements of the cosmic microwave background (CMB) anisotropies by the \textit{Planck} satellite, are consistent with a primordial power spectrum (PPS) of curvature perturbations predicted by cosmic inflation, which is adiabatic, almost Gaussian, and nearly scale-invariant~\cite{Planck:2018vyg,Akrami:2018odb,Planck:2019izv}. The simple power-law parametrisation, $\mathcal{P}_\zeta(k) \propto k^{n_\mathrm{s}-1}$, aligns well with the predictions of single-field slow-roll inflation~\cite{Starobinsky:1980te,Guth:1980zm,Linde:1981mu,Mukhanov:1981xt}. However, from a theoretical perspective, the assumption of a featureless power spectrum over all scales is likely an oversimplification. The inflationary period may have involved complex dynamics, leaving, for instance, distinct oscillatory signatures in the PPS~\cite{Chluba:2015bqa,2022arXiv220308128A}.

These features typically manifest as oscillatory patterns in the PPS, with frequencies that are either linear ($k$) or logarithmic ($\ln k$) in wavenumber space, breaking scale invariance locally or globally. While extensive searches have been performed using CMB temperature and polarisation power spectra~\cite{Adams:2001vc,Peiris:2003ff,Mukherjee:2003ag,Martin:2003sg,Covi:2006ci,Hamann:2007pa,Meerburg:2011gd,Flauger:2009ab,Aich:2011qv,Chen:2012ja,Peiris:2013opa,Planck:2013jfk,Planck:2013wtn,Meerburg:2013dla,Benetti:2013cja,Miranda:2013wxa,Easther:2013kla,Chen:2014joa,Achucarro:2014msa,Hazra:2014goa,Hazra:2014jwa,Hu:2014hra,Fergusson:2014tza,Planck:2015zfm,Ade:2015lrj,Gruppuso:2015xqa,Hazra:2016fkm,Torrado:2016sls,Akrami:2018odb,Handley:2019fll,Zeng:2018ufm,Planck:2019izv,Canas-Herrera:2020mme,Braglia:2021ckn,Braglia:2021sun,Braglia:2021rej,Naik:2022mxn,Hamann:2021eyw,Antony:2024vrx,Raffaelli:2025kew,Raffaelli:2026bhk} and, more recently, large-scale structure (LSS) clustering~\cite{Beutler:2019ojk,Ballardini:2022wzu,Mergulhao:2023ukp,Calderon:2025xod}, no statistically compelling detection has been claimed to date. However, the search for primordial physics is not limited to two-point correlators. Many physical mechanisms generating oscillatory features in the power spectrum simultaneously induce specific, correlated non-Gaussian signatures in the bispectrum~\cite{Chen:2006xjb,Flauger:2009ab,Flauger:2010ja,Achucarro:2010da,Adshead:2011jq}. For instance, sharp or resonance feature scenarios predict a non-trivial scale dependence of the non-Gaussianity shape, where the bispectrum amplitude oscillates with a phase and frequency strictly related to those appearing in the power spectrum, potentially accompanied by a running of the amplitude. Consequently, a joint analysis of the power spectrum and bispectrum represents a significantly more powerful probe than either statistic alone, lifting degeneracies and potentially enhancing the signal-to-noise ratio (SNR) of a discovery; see Refs.~\cite{Fergusson:2014hya,Fergusson:2014tza,Meerburg:2015owa,Akrami:2018odb} for studies combining power spectrum and bispectrum searches for primordial oscillatory features in CMB measurements.

In this context, future radio surveys mapping the neutral hydrogen (HI) distribution offer a distinct advantage over current optical galaxy surveys and CMB experiments; see Refs.~\cite{Chen:2016zuu,Xu:2016kwz,Meerburg:2016zdz,Ballardini:2017qwq,Bacon:2018dui} for previous forecast studies. By using both single-dish intensity mapping and interferometric observations, radio surveys can access an unprecedented comoving volume and a wide range of scales, extending deep into the dark ages or the high-redshift matter-dominated era. This vast observational window allows for the sampling of a large number of modes, which is critical for resolving high-frequency oscillatory features that are otherwise damped or smoothed out in projected CMB observables. Beyond the power spectrum, the bispectrum provides significant information by probing mode coupling and non-Gaussian signatures in LSS. Recent optical survey analyses have demonstrated its constraining power, both alone and combined with the power spectrum, for cosmological parameters and primordial non-Gaussianity (PNG)~\cite{Castorina:2019wmr,Mueller:2021jbt,Philcox:2021kcw,Ivanov:2021kcd,DAmico:2022gki,Cabass:2022wjy,Cabass:2022ymb,Sugiyama:2023tes,Ivanov:2023qzb,Cagliari:2025rqe}. Radio surveys provide a promising extension of these studies, particularly for bispectrum analyses. Previous HI intensity mapping (IM) forecasts have shown that interferometric surveys can be especially powerful for bispectrum measurements, outperforming single-dish surveys for PNG constraints in several cases~\cite{Karagiannis:2019jjx,Karagiannis:2020dpq,Karagiannis:2022ylq}. These results motivate the joint power spectrum and bispectrum analysis of future radio surveys adopted in this work, applied to primordial feature models.

In this work, we present a systematic forecast analysis for the detection of primordial oscillatory features, focusing on the synergy between the power spectrum and the bispectrum. Unlike previous studies that often treat these statistics independently or focus solely on the CMB, we construct a combined analysis for future radio surveys, considering both single-dish and interferometric configurations. We examine specific theoretical models -- including linear and logarithmic oscillations -- where the relationship between the two- and three-point functions is theoretically determined, allowing for a consistent combined search. To ensure a robust observational baseline, we complement the radio forecasts with CMB information, included at the level of the angular power spectrum. This approach aims to quantify the added value of the LSS bispectrum in constraining the amplitude and frequency of primordial features and to assess the capability of next-generation radio experiments to test inflationary physics beyond the standard slow-roll predictions.

The paper is organised as follows. In~\cref{sec:theory}, we introduce the power spectrum and bispectrum templates used in this work and describe the $\mathrm{HI}$ intensity mapping modelling for both in redshift space. In~\cref{sec:method}, we discuss our forecast methodology for HI intensity mapping and CMB observables. In~\cref{sec:results}, we present our results and conclude in~\cref{sec:conclusion}. We provide additional checks in the appendices: \cref{app:Amplx10} reports forecasts for a feature amplitude ten times larger than the fiducial value, while \cref{app:sn_triangles} shows the frequency dependence of the bispectrum SNR across triangle configurations.

\section{Theoretical model} \label{sec:theory}
In this section, we summarise the theoretical link between primordial inflationary fluctuations and late-time LSS observables. We introduce the primordial curvature power spectrum and bispectrum (including oscillatory feature templates), and show how they map to the linear matter power spectrum and the linearly-propagated primordial contribution to the matter bispectrum. Subsequent sections will incorporate biasing, redshift-space distortions, and infrared (IR) resummation for the HI IM observables.

\subsection{Primordial power spectrum, bispectrum and features} \label{sec:theory_templates}
We define the (comoving) curvature perturbation in Fourier space, $\zeta(\bk)$, through
\begin{equation}
    \langle \zeta(\bk)\,\zeta(\bk')\rangle = (2\pi)^3 \delta_{\rm D}(\bk+\bk')\,P_\zeta(k) \,,
\end{equation}
where statistical isotropy implies that $P_\zeta$ depends only on $k\equiv |\bk|$. It is convenient to introduce the dimensionless power spectrum
\begin{equation} \label{eq:PPS}
    \Delta^2_{\zeta,0}(k)\equiv \frac{k^3}{2\pi^2}P_{\zeta,0}(k) = A_\mathrm{s}\left(\frac{k}{k_*}\right)^{n_\mathrm{s}-1} \,,
\end{equation}
with pivot scale $k_* = 0.05\,\mathrm{Mpc}^{-1}$, amplitude $A_\mathrm{s}$, and scalar spectral index $n_\mathrm{s}$.
On super-horizon scales during matter domination, the Bardeen potential is related to $\zeta$ by $\Phi = \tfrac{3}{5}\,\zeta$, so that $P_\Phi(k)=\tfrac{9}{25}\,P_\zeta(k)$.

The linear matter power spectrum is obtained via the Poisson equation,
\begin{equation}
    P_{\rm lin}(k,z) = \mathcal{M}^2(k,z)\,P_\Phi(k) \,,
\end{equation}
where
\begin{equation}
    \mathcal{M}(k,z)=\frac{2k^2c^2}{3\Omega_\mathrm{m}H_{0}^2}\,T(k)\,D(z) \,.
\end{equation}
Here $T(k)$ is the matter transfer function (normalised to $T\to 1$ as $k\to 0$) and $D(z)$ is the linear growth factor (with the usual normalisation, i.e.\ $D(0)=1$).

PNG is encoded in higher-order correlators, with the leading contribution given by the bispectrum. We define the primordial bispectrum of $\Phi$ as
\begin{equation}
    \langle \Phi(\bk_1)\Phi(\bk_2)\Phi(\bk_3)\rangle = (2\pi)^3\delta_{\rm D}(\bk_1+\bk_2+\bk_3)\,B_{\Phi}(k_1,k_2,k_3) \,,
\end{equation}
where the Dirac delta imposes the triangle condition. We write
\begin{equation}
    B_\Phi(k_1,k_2,k_3)= f_{\rm NL}\,F(k_1,k_2,k_3) \,,
\end{equation}
where $f_{\rm NL}$ is a dimensionless amplitude and $F$ a shape function specifying the triangle dependence~\cite{Komatsu2009}.
The linearly-propagated primordial contribution to the (matter) bispectrum is then
\begin{equation} \label{eq:BI_def}
    B_{\rm I}(k_1,k_2,k_3;z)=\mathcal{M}(k_1,z)\mathcal{M}(k_2,z)\mathcal{M}(k_3,z)\,B_{\Phi}(k_1,k_2,k_3) \,.
\end{equation}
In real space, statistical isotropy implies $B_{\rm I}$ depends only on $(k_1,k_2,k_3)$; in redshift space, it acquires an additional dependence on the angles with respect to the line of sight.

Non-trivial inflationary dynamics can imprint oscillatory features in primordial correlators. Phenomenologically, these signals are commonly grouped into:
(i) \emph{sharp feature} \cite{Starobinsky:1992ts,Adams:2001vc,Chen:2006xjb,Chen:2008wn,Achucarro:2010da,Adshead:2011jq}, generated by brief violations of slow-roll (e.g.\ steps or sudden turns), which typically produce oscillations approximately linear in $k$; and
(ii) \emph{resonant feature} \cite{Chen:2008wn,Flauger:2009ab,Flauger:2010ja,Chen:2010bka}, sourced by persistent periodic modulations of the background, leading to oscillations approximately linear in $\ln k$.

We model oscillations in the PPS as a fractional modulation around the featureless baseline $P_{\zeta,0}$,
\begin{equation} \label{eq:feature_PS}
    P^X_\zeta(k) = P_{\zeta,0}(k)\left[1 + \delta P^X(k)\right] \,,
\end{equation}
with
\begin{equation} \label{eq:feat_PS_w}
    \delta P^X(k) = \mathcal{A}_X \sin\left[\omega_X\,\Xi_X(k)+ 2\pi\phi_X^P\right] \,,
\end{equation}
where $X\in\{{\rm lin,log}\}$, $\Xi_{\rm lin}(k)=k/k_*$ and $\Xi_{\rm log}(k)=\ln(k/k_*)$. The parameters $\mathcal{A}_X$, $\omega_X$, and $\phi_X^P$ denote the amplitude, dimensionless frequency, and normalised phase.

The associated oscillatory primordial bispectrum template is taken as \cite{Chen:2006xjb,Chen:2008wn,Flauger:2010ja,Chen:2010bka}
\begin{equation} \label{eq:Bk_osc}
    B_{\Phi}^X(k_1,k_2,k_3) = f_{\rm NL}^X\, \frac{6A^2}{k_1^2k_2^2k_3^2}\,
    \sin\!\left[\omega_X\,O_X(K)+2\pi\phi_X^B\right] \!,
\end{equation}
where $K\equiv k_1+k_2+k_3$, $O_{\rm lin}(K)=K/k_*$ and $O_{\rm log}(K)=\ln(K/k_*)$.
The normalisation $A$ is defined by writing the featureless potential power spectrum as
$P_\Phi(k)=A\,k^{n_\mathrm{s}-4}$ (equivalently $P_\Phi\propto k^{n_\mathrm{s}-4}$).%
\footnote{Using $\Phi=3\zeta/5$ and \cref{eq:PPS}, one finds
$A={18\pi^2}A_\mathrm{s}k_*^{1-n_\mathrm{s}}/25$ for the pure power-law normalisation.}
Each phenomenological feature model is thus specified by five parameters 
$\{\mathcal{A}_X,\omega_X,\phi_X^P,f_{\rm NL}^X,\phi_X^B\}$.
In this agnostic parametrisation, we correlate the PPS and the primordial bispectrum only through a common oscillation frequency $\omega_X$, while allowing their amplitudes and phases to vary independently.

In addition to the above phenomenological parametrization, we also consider a model with a sharp feature generated by a step in the potential for the inflaton field. In this case, the correlated PPS and primordial bispectrum can be consistently calculated at first-order generalised slow-roll (GSR) within the same underlying model; see Refs.~\cite{Stewart:2001cd,Dvorkin:2009ne,Bartolo:2013exa}.

At first order, the curvature power spectrum can be written as
\begin{equation}\label{eq:gsr_ps}
    \ln \Delta^2_\zeta(k) = \ln \Delta_{\zeta,0}^2(k) + \frac{\mathcal{C}}{3}\,
    \mathcal{D}\!\left(\frac{k\eta_f}{x_d}\right)\,
    W'(k\eta_f) \,,
\end{equation}
where $\Delta_{\zeta,0}^2$ is the featureless PPS, $\eta_f$ is the conformal time at which the feature is crossed, $x_d$ controls the width (sharpness) of the transition, and $\mathcal{C}$ sets the feature amplitude. The damping envelope is
\begin{equation}
    \mathcal{D}(y) = \frac{y}{\sinh y} \,,
\end{equation}
and the window function is
\begin{equation}
    W(x)=\frac{3\sin(2x)}{2x^3}-\frac{3\cos(2x)}{x^2}-\frac{3\sin(2x)}{2x}\,,
    \qquad W'(x)\equiv \frac{\mathrm{d}W}{\mathrm{d}\ln x}=x\frac{\mathrm{d}W}{\mathrm{d}x} \,,
\end{equation}
with the explicit expression
\begin{equation}
    W'(x)=\left(-3+\frac{9}{x^2}\right)\cos(2x)+\left(15-\frac{9}{x^2}\right)\frac{\sin(2x)}{2x} \,.
\end{equation}
In analogy with~\cref{eq:feature_PS}, we can define
\begin{equation} \label{eqn:feat_step_w}
    \delta P^\mathrm{step}(k) = \exp\left[\frac{\mathcal{C}}{3}\mathcal{D}\!\left(\frac{k\eta_f}{x_d}\right)W'(k\eta_f)\right] - 1 \,.
\end{equation}

\begin{figure}
    \centering
    \includegraphics[width=\linewidth]{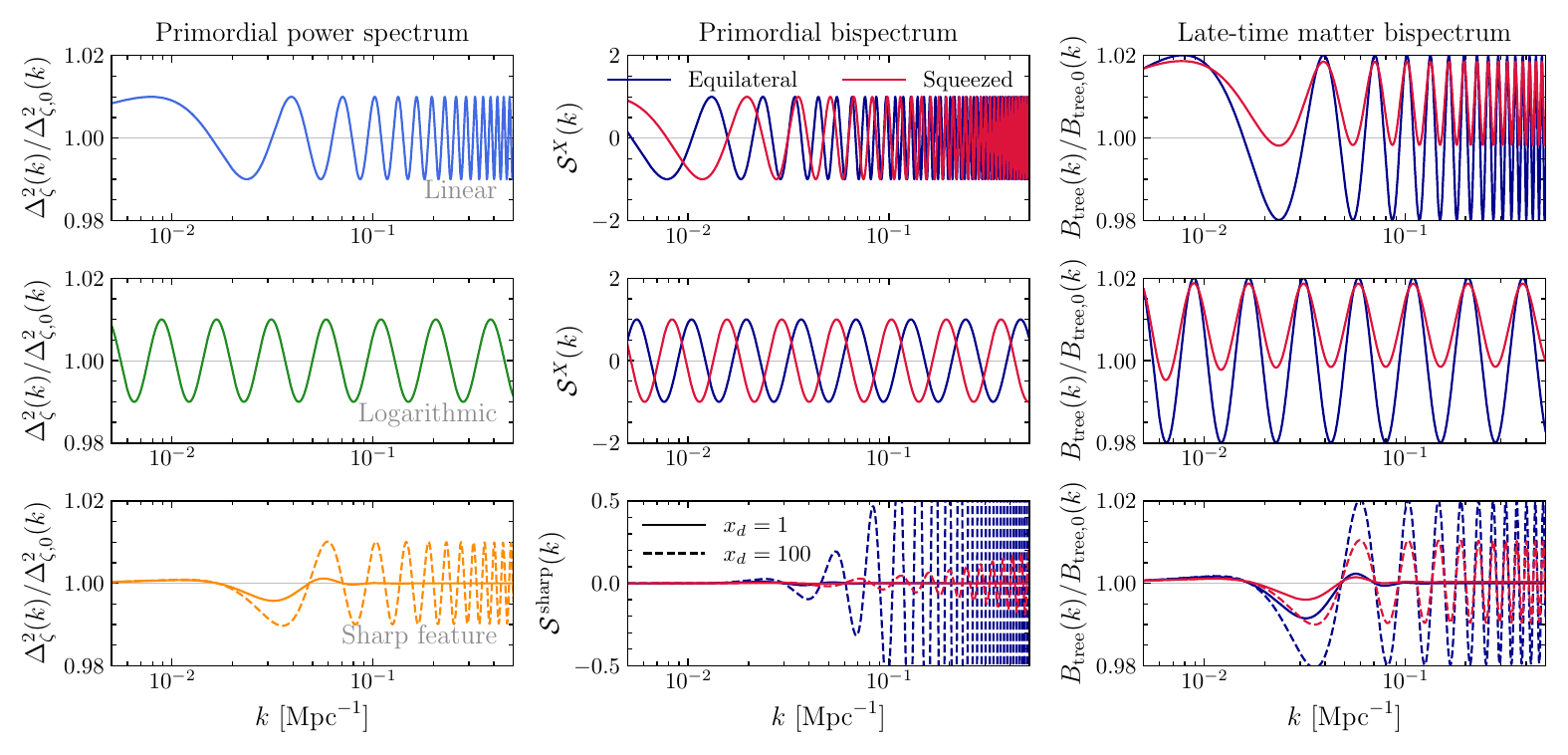}
    \caption{Primordial feature templates for the three models: linear oscillations (top row),
    logarithmic oscillations (middle row), and sharp feature (bottom row).
    \textit{Left column}: power spectrum ratio $\Delta^2_\zeta(k)/\Delta^2_{\zeta,0}(k)$,
    from \cref{eq:feature_PS,eq:gsr_ps}.
    \textit{Middle column}: bispectrum shape functions $\mathcal{S}^X$ and $\mathcal{S}^{\rm sharp}$
    from \cref{eq:linlog_source,eq:gsr_source} respectively, at equilateral (blue) and squeezed (red) configurations.
    \textit{Right column}: tree-level matter bispectrum ratio $B_{\rm tree}(k)/B_{{\rm tree},0}(k)$, where $B_{{\rm tree},0}$ is computed with the featureless PPS. For the sharp-feature model, solid and dashed curves correspond to $x_d=1$ and $x_d=100$. All curves use the fiducial parameter values of each model.}
    \label{fig:fid_templates}
\end{figure}
    
The primordial curvature bispectrum generated by the same sharp dynamics \cite{Adshead:2011bw,Adshead:2011jq,Bartolo:2013exa} can be written as
\begin{align} \label{eq:gsr_bzeta}
    B_\zeta(k_1,k_2,k_3)
    &=\frac{(2\pi)^4}{(k_1k_2k_3)^3}\,
    \frac{A_\mathrm{s}^{2}}{4}
    \left(\frac{k_1k_2k_3}{k_*^3}\right)^{\frac{n_\mathrm{s}-1}{2}}
    \nonumber\\
    &\hspace{1.2cm}\times
    \left[
    -I_0(K)\,k_1k_2k_3
    -I_1(K)\!\!\sum_{i\neq j}\!k_i^2k_j
    +I_2(K)\,K\,(k_1^2+k_2^2+k_3^2)
    \right]\!,
\end{align}
where 
\begin{equation}\label{eq:In_def}
    I_n(K) = \frac{\mathcal{C}}{2}\,
    \mathcal{D}\!\left(\frac{K\eta_f}{2x_d}\right)\,
    X_n(K\eta_f)\,.
\end{equation}
The basis functions are
\begin{align}
    X_0(x) &= -\frac{(x^4-9x^2+54)\cos x}{x^2}
    +\frac{(2x^4-27x^2+54)\sin x}{x^3}\,,
    \\
    X_1(x) &= \frac{3(x^2-6)\cos x}{x^2}
    +\frac{(x^2-6)(x^2-3)\sin x}{x^3}\,,
    \\
    X_2(x) &= -\frac{3(x^2-9)\cos x}{x^2}
    +\frac{(4x^2-9)\sin x}{x^3}\,.
\end{align}
If needed, one can convert to the potential bispectrum as $B_\Phi = \left(\tfrac35\right)^3 B_\zeta$.

This template makes explicit two generic properties of sharp features.
First, oscillations are exponentially damped at large $k$ (or large $K$) by the envelope $\mathcal{D}$, with $x_d$ controlling how rapidly the signal decays towards small scales.
Second, $\eta_f$ fixes the oscillation phase coherently across correlators, effectively tying together the characteristic frequency and phase of the oscillations in both the power spectrum and bispectrum. In practice, the combination of the IR behavior of the $X_n$ functions and the UV damping from $\mathcal{D}$ localizes the signal into a wave-packet in momentum space.

For later intuition, the effective bispectrum amplitude in this class of models is expected to scale approximately as
$f_{\rm NL}^{\rm eff}\propto \mathcal{C}\,x_d^{2}$ (up to an $\mathcal{O}(1)$ factor) \cite{Chen:2011zf,Bartolo:2013exa}. 
In particular, around the peak of the oscillatory wave-packet (where the envelope $\mathcal{D}$ is maximal), the effective bispectrum amplitude can readily exceed unity and may reach $f_{\rm NL}^{\rm eff}\sim 10$--$100$ for sufficiently sharp transitions (large $x_d$) and moderate feature amplitudes $\mathcal{C}$, while remaining strongly scale dependent and damped away from the peak. 

To compare different primordial bispectrum templates on equal footing, we factor out the universal dimensional scaling and define an amplitude-weighted shape function $\mathcal{S}(k_1,k_2,k_3)$ through
\begin{equation}
    B_\zeta^X(k_1,k_2,k_3) = \frac{18}{5}\frac{(2\pi^2)^2\Delta_\zeta^4}{(k_1k_2k_3)^2}\,\mathcal{S}^X(k_1,k_2,k_3) \,.
\end{equation}
This definition removes the common scale dependence associated with the nearly scale-invariant power spectrum and isolates the physical momentum dependence of the bispectrum. For the phenomenological linear and logarithmic templates,
\begin{equation} \label{eq:linlog_source}
    \mathcal{S}^X(k_1,k_2,k_3) = f_\mathrm{NL} \sin\left[\omega_XO_X(K) + 2\pi\phi_X^B\right] \!,
\end{equation}
while for the step model the same quantity contains the full amplitude generated by the source functions: 
\begin{equation} \label{eq:gsr_source}
    \mathcal{S}^{\rm sharp}(k_1,k_2,k_3) = \frac{5}{18} \left[-I_0(K)-I_1(K)\frac{\sum_{i\neq j}\!k_i^2k_j}{k_1k_2k_3}
    +I_2(K)\,K\,\frac{k_1^2+k_2^2+k_3^2}{k_1k_2k_3}\right] \!.
\end{equation}
The amplitude-weighted shape function $\mathcal{S}(k_1,k_2,k_3)$ therefore provides a direct comparison of the phase, frequency, envelope and effective non-Gaussian amplitude of the different models, where the normalisation removes the common dimensional prefactor of the primordial curvature bispectrum. 

In~\cref{fig:fid_templates}, we show the fiducial PPS, as a ratio with respect to the featureless spectrum according to~\cref{eq:feat_PS_w,eqn:feat_step_w}, and the primordial bispectra, as amplitude-weighted shape functions according to~\cref{eq:linlog_source,eq:gsr_source}.

\subsection{HI intensity mapping power spectrum and bispectrum in redshift space} \label{sec:HI_model}
We adopt a tree-level perturbation-theory description for the HI IM power spectrum and bispectrum, supplemented by a phenomenological Finger-of-God (FoG) damping term, which is adequate on the scales considered here \cite{Gil-Marin:2014sta,Lazanu2015b,Hashimoto:2017klo,Chan2017,Oddo:2019run,Agarwal:2020lov,MoradinezhadDizgah2020}. 
In the presence of primordial features, the redshift-space HI power spectrum and bispectrum read
\begin{align}
    P_{\rm HI}(\bk,z) &= \bar{T}_b(z)^2\left[D_\text{FOG}^P(\bk,z)Z_1(\bk,z)^2P_{\rm lin}^X(k,z)+P_{\veps}(z)\right]+P_{\rm N}(\bk,z) \,, \label{eq:Pgs}\\ 
    B_{\rm HI}(\bk_1,\bk_2,\bk_3,z) &= \bar{T}_b(z)^3 \Bigg\{D_\text{FOG}^B(\bk_1,\bk_2,\bk_3,z)\bigg[Z_1(\bk_1,z)Z_1(\bk_2,z)Z_1(\bk_3,z)B_{\rm I}^X(k_1,k_2,k_3,z) \nonumber \\ 
   &~~+\left(2Z_1(\bk_1,z)Z_1(\bk_2,z)Z_2(\bk_1,\bk_2,z)P_{\rm lin}^X(k_1,z)P_{\rm lin}^X(k_2,z)+2~ \text{perm}\right)\bigg] \nonumber \\
   &~~+2P_{\veps\veps_{\delta}}(z)\Big[Z_1(\bk_1,z)P_{\rm lin}^X(k_1,z)+2~ \text{perm}\Big]+B_{\veps}(z)\Bigg\} \,. \label{eq:Bgs} 
  \end{align}
Here $P_{\rm lin}^X(k,z) = P_\mathrm{lin}(k,z)\left[1 + \delta P^X(k)\right]$ is the linear matter power spectrum including the oscillatory features, and $B_{\rm I}^X$ is the linearly-propagated primordial contribution to the matter bispectrum, as defined in~\cref{sec:theory_templates}.
The instrumental noise $P_{\rm N}$ depends on the experimental setup and observing mode (see~\cref{sec:method}). The stochastic contributions $P_\epsilon$, $P_{\epsilon\epsilon_\delta}$, and $B_\epsilon$ are treated as scale independent at this order.

In~\cref{fig:fid_templates}, we show the tree-level matter bispectrum on the right column, as a ratio with respect to the featureless one. 
In the equilateral limit ($k_1=k_2=k_3$), all three modes share the same physical horizon and co-evolve at a single scale, preserving symmetric oscillations around the baseline. Conversely, the squeezed limit ($k_3 \to 0, \, k_1 \approx k_2$) maps how a long-wavelength background mode modulates rapid, short-wavelength fluctuations. For the linear and logarithmic templates, primordial oscillations remain active throughout the inflationary phase, meaning the long mode $q$ is itself oscillatory and acts as an undulating background that distorts the short-mode envelope, sourcing non-symmetric wave-beats through late-time gravitational coupling $\propto P(k)P(q)$. In contrast, for sharp features (the step model), the localized transition only affects small-scale modes that are sub-horizon at the time of the event. Large-scale modes ($q$) exit the horizon prior to the feature triggering, leaving the long-mode power spectrum entirely unmodulated because of the damping envelope.

The mean HI brightness temperature is
\begin{equation}
    \bar{T}_b(z)=188\,\Omega_{\rm HI}(z)\,h\,(1+z)^2\frac{H_0}{H(z)}~\,{\rm mK} \,,
\end{equation}
where we model the HI density evolution as $\Omega_{\rm HI}(z)=4\times 10^{-4}(1+z)^{0.6}$ \cite{Battye2013}. 

The redshift-space kernels are~\cite{Baldauf2011,Tellarini2016,Karagiannis2018}
\begin{align}
   &Z_1(\bk_i,z)=b_1(z)+f\mu_i^2\,, \label{eq:Z1}\\
   &Z_2(\bk_i,\bk_j,z)=b_1(z)F_2(\bk_i,\bk_j)+f(z)\mu_{ij}^2G_2(\bk_i,\bk_j)+\frac{b_2(z)}{2} \nonumber \\
   & +\frac{b_{s^2}(z)}{2}S_2(\bk_i,\bk_j)+\frac{f(z)\mu_{ij}k_{ij}}{2}\left[\frac{\mu_i}{k_i}Z_1(\bk_j,z)+\frac{\mu_j}{k_j}Z_1(\bk_i,z)\right] \!, \label{eq:Z2}
\end{align}  
where $f(z)\equiv \mathrm{d}\ln D/\mathrm{d}\ln a$ is the linear growth rate, $\mu_i=\hat\bk_i\cdot\hat{\bm{z}}$ with $\hat{\bm z}$ the line-of-sight vector, $\mu_{ij}=(\mu_ik_i+\mu_jk_j)/k_{ij}$ and $k_{ij}^2=(\bk_i+\bk_j)^2$. The kernels $F_2$ and $G_2$ are the second-order symmetric SPT kernels \cite{Bernardeau2002}, $S_2(\bk_1,\bk_2) = (\hat\bk_1\cdot\hat\bk_2)^2-1/3$ is the tidal kernel \cite{McDonald2009,Baldauf2012}. 

We include a Gaussian FoG damping~\cite{Peacock1994,Ballinger1996}
\begin{align}
    D_\text{FOG}^P(\bk) &= \exp\left[-\big(k\mu\sigma_P\big)^2\right] \!, \\
    D_\text{FOG}^B(\bk_1,\bk_2,\bk_3) &= \exp\left[-\left(k_1^2\mu_1^2+k_2^2\mu_2^2+k_3^2\mu_3^2\right)\sigma_B^2\right] \!,
\end{align}
with fiducial values $\sigma_P=\sigma_B=2\,\sigma_v$, where the pairwise peculiar velocity dispersion along the line of sight is estimated as
\begin{equation}
    \sigma_v^2(z)=\frac{f(z)^2}{6\pi^2}\int_0^\infty {\rm d}k\,P_{\rm lin}(k,z) \,.
\end{equation}

Oscillatory primordial bispectra can induce scale-dependent bias signatures, analogous to the local PNG case~\cite{Dalal2008,Slosar2008,Matarrese2008}, entering at linear order in the galaxy power spectrum and bispectrum. However, Ref.~\cite{Cabass:2018roz} demonstrated that these corrections exhibit oscillatory behaviour with an envelope similar to that of equilateral PNG~\cite{Schmidt2010,Scoccimarro2011,Schmidt2012,Assassi2015}, and with amplitudes too small to be detectable by upcoming surveys. We therefore neglect these non-Gaussian scale-dependent bias contributions in the redshift kernels defined above.

We model the HI bias parameters using the halo-model framework~\cite{Seljak2000,Peacock2000,Scoccimarro2000}, where the halo mass function used is from Ref.~\cite{Tinker2010} and the halo occupation distribution from Ref.~\cite{Castorina2016}. The HI bias coefficients follow from mass-weighted integrals over the halo bias~\cite{Santos:2015gra,Karagiannis:2020dpq,Karagiannis:2022ylq}, with linear halo bias from~\cite{Tinker2010}, quadratic bias from peak-background split~\cite{Lazeyras2016,Karagiannis:2019jjx}, and tidal bias via $b_{s^2}=-4(b_1-1)/7$~\cite{Baldauf2012}. Stochastic terms are set to their Poisson predictions~\cite{Schmidt2015,Desjacques2016}, with the shot-noise power spectrum given in Ref.~\cite{Castorina2016}, though shot-noise contributions are subdominant in HI IM compared to instrumental noise~\cite{Santos:2015gra,Bull2015}. We also account for the Alcock--Paczynski (AP) effect~\cite{Alcock1979}, which induces anisotropic distortions when the fiducial cosmology used to convert angles and redshifts into comoving distances differs from the true one~\cite{Seo2003,Song2015}. This induces an artificial anisotropic distortion of the inferred density field, modifying both the amplitude and shape of the power spectrum and bispectrum (see e.g. Ref.~\cite{Karagiannis:2022ylq} for details). 

Large-scale bulk flows generate infrared-enhanced perturbative contributions that smear oscillatory features in Fourier space, including the BAO and any superimposed primordial oscillations~\cite{Crocce:2007dt,Creminelli:2013poa,Senatore:2014via,Baldauf:2015xfa,Blas:2016sfa}.
These effects are important already at tree level and are most conveniently incorporated through IR resummation~\cite{Baldauf:2015xfa,Blas:2016sfa,Ivanov:2018gjr}.
In practice, we implement this by replacing the linear matter power spectrum entering~\cref{eq:Pgs,eq:Bgs} with an IR-resummed version, following the TSPT-based treatment~\cite{Blas:2015qsi} of primordial oscillatory features~\cite{Blas:2016sfa,Vasudevan:2019ewf,Beutler:2019ojk,Euclid:2023shr,Ballardini:2024dto,Chen:2024pyp,Goldstein:2025eyj}:

\begin{align}
    P_{\rm lin}^X(k,z)\to P^{\rm IR}_{\rm X}(\bk,z) = &P^{\rm nw}(k,z) + P^{\rm w}_{\rm BAO}(k,z)\,\mathrm{e}^{-k^2\Sigma^2_{\rm BAO}(z)\,g(\mu,z)} \notag\\
    &+ P^{\rm nw}(k,z)\delta P^X(k)\,\mathrm{e}^{-k^2\Sigma^2_{\rm X}(z)\,g(\mu,z)}(k,z)\mathcal{F}_{\rm X}(\bk,z) \,,
\end{align}
where $g(\mu,z)\equiv 1+f(z)\mu^2[2+f(z)]$~\cite{Eisenstein:2006nk}. For linear oscillations we take $\mathcal{F}_{\rm lin}(\bk,z)=1$, while for logarithmic oscillations, IR resummation induces an additional phase shift that can be captured by $\mathcal{F}_{\rm log}(\bk,z)=\cos\!\big[k^2\Sigma^2_{\rm log}(k,z)\,g(\mu,z)\big]$. We extract the part of the linear matter power spectrum that contains the BAO wiggles, i.e. $P^{\rm w}_{\rm BAO}$, by using a Savitzky-Golay filter~\cite{Boyle:2017lzt}.
We then define $P^{\rm nw}=P_{\rm lin}-P^{\rm w}_{\rm BAO}$. The damping functions are
\begin{align}
    \Sigma_{\rm BAO}^2(z) &\equiv \int_0^{k_{\rm S}}\frac{{\rm d}q}{6\pi^2} P^{\rm nw}(q,z)\left[1-j_0(qr_{\rm s})+2j_2(qr_{\rm s})\right] \!, \label{eq:sigma_bao} \\
    \Sigma_{\rm lin}^2(z) &\equiv \int_0^{k_{\rm S}}\frac{{\rm d}q}{6\pi^2} P^{\rm nw}(q,z)\left[1-j_0\!\left(\frac{q\omega_{\rm lin}}{k_*}\right)+2j_2\!\left(\frac{q\omega_{\rm lin}}{k_*}\right)\right] \!, \label{eq:sigma_lin} \\
    \Sigma^2_{\rm log}(k,z) &\equiv \int_0^{k_{\rm S}}\frac{{\rm d}q}{4\pi^2} P^{\rm nw}(q,z)\int_{-1}^1{\rm d}\mu\,\mu^2\left\{1-\cos\!\left[\omega_{\rm log}\ln\!\left(1-\frac{q\mu}{k}\right)\right]\right\} \!, \label{eq:sigma_log} 
\end{align} 
where $j_n$ are spherical Bessel functions, $r_{\rm s} \simeq 147\,{\rm Mpc}$ is the BAO sound horizon~\cite{Planck:2018vyg}, and $k_{\rm S}$ is the resummation separation scale. Following Refs.~\cite{Baldauf:2015xfa,Vasudevan:2019ewf,Ballardini:2024dto}, we set $k_{\rm S}=0.2\,h\,{\rm Mpc}^{-1}$. 

Long-wavelength displacements also damp oscillatory primordial bispectra. A systematic IR resummation for oscillatory PNG shows that, at leading order, the wiggly component of the primordial bispectrum is multiplicatively suppressed by an exponential factor, with a damping scale that depends on the triangle shape and on the oscillation frequency. In particular, the damping reduces to the corresponding power-spectrum damping in the squeezed limit~\cite{Blas:2016sfa,Vasudevan:2019ewf}.
In our baseline model we account for this effect by applying a Gaussian IR suppression to the linearly-propagated primordial oscillatory contribution,
\begin{equation} \label{eq:BI_IR}
    B_{\rm I}(k_1,k_2,k_3;z)\;\longrightarrow\;
    B_{\rm I}^{\rm IR}(\bk_1,\bk_2,\bk_3;z)
    = B_{\rm I}(k_1,k_2,k_3;z)\exp\big[{-\Sigma_{\rm B}^2(k_1,k_2,k_3;z)}\big] \,,
\end{equation}
where $\Sigma_{\rm B}^2(k_1,k_2,k_3;z) = \sum_{i<j}^3 k_i k_j\,J(\,\hat{\bk}_i\!\cdot\!\hat{\bk}_j\,,z)$.
As a simple high-frequency approximation (phase-averaged limit), we retain only the leading term of the full leading-order kernel,
\begin{align}
    J(x,z) \simeq -x\,\int_0^{k_{\rm S}}\frac{\mathrm{d}q}{6\pi^2}\,P_{\rm lin}(q,z) \,.
\end{align}
This prescription corresponds to treating long-wavelength bulk flows in the Zel'dovich approximation, i.e.\ using linear (Gaussian) displacement fields to model the IR smoothing of oscillatory contributions.  This treatment has recently been validated against N-body simulations with non-Gaussian initial conditions for oscillatory squeezed bispectra~\cite{Goldstein:2025eyj}, which also show that the oscillatory signal is strongly damped on small scales ($k\gtrsim 0.3\,h\,{\rm Mpc}^{-1}$ at $z=0$). While this is expected to be a good estimate for relatively high oscillation frequencies, in TSPT the full leading-order result features a non-trivial dependence on the triangle shape through spherical Bessel functions; see Ref.~\cite{Vasudevan:2019ewf} for the complete expression in the resonant case.

\section{Forecast methodology} \label{sec:method}
In this section we describe the forecasting methodology adopted in our analysis, together with the main characteristics of the datasets considered. We employ an information matrix approach to quantify the constraining power of future cosmological observations. 
Our parameter space comprises the five parameters characterising the primordial feature model, alongside the standard cosmological parameters. 
In addition, we vary the relevant 21cm nuisance parameters entering the IM signal, including the HI bias parameters. 
Radio and CMB observations are treated as statistically independent probes; under this assumption, their information matrices are combined by direct summation. 

The fiducial cosmology is given by $\omega_{\rm b}=0.02237$, $\omega_{\rm c}=0.12$, $h=0.6736$, $\ln(10^{10}A_{\rm s})=3.044$, $n_{\rm s}=0.9649$, and $\sum m_\nu=0.06,\mathrm{eV}$, consistent with the \textit{Planck} 2018 baseline results \cite{Planck:2018vyg}.
For the primordial feature sector, we adopt the fiducial values ${\cal A}_X=0.01$, $\omega_X = 10$, $\phi_X^P = 0$, $\fnl^X = 1$, and $\phi_X^B = 0$ for the sharp and resonant feature models. For the sharp-feature parametrisation defined in~\cref{eq:gsr_ps,eq:gsr_bzeta}, we assume ${\cal C} = 0.01$, $\eta_f = 50\,h\,{\rm Mpc}^{-1}$, and $x_d = 1,\,100$. 
The PPS and primordial bispectrum for the feature models and fiducial values considered are shown in~\cref{fig:fid_templates}. The figure also shows the corresponding late-time tree-level matter bispectrum, illustrating that the oscillatory structure of the primordial power spectrum is inherited to the gravitational late-time bispectrum.

\subsection{HI intensity mapping forecasting methodology} \label{sec:method_HI}
To forecast the potential of upcoming HI IM surveys in constraining primordial features, we utilise the information matrix methodology. The information matrix at redshift $z$ for the power spectrum is
\begin{equation} \label{eq:fisherPs}
    \mathcal{I}_{\alpha\beta}^{\rm HI, \,P}(z)=\sum_{k,k'}\int_{-1}^1 \frac{\mathrm{d}\mu}{2} \frac{\partial P_{\rm HI}(k,\mu,z)}{\partial \theta_{\alpha}}\,\left[C^{\rm P}_{kk'}\right]^{-1}\,\frac{\partial P_{\rm HI}(k',\mu,z)}{\partial \theta_{\beta}} \,,
\end{equation}
while for the bispectrum
\begin{equation} \label{eq:fisherBs}
    \mathcal{I}_{\alpha\beta}^{\rm HI,\,B}(z)=\int {\rm d}^3\bk_i \,\frac{\partial B_{\rm HI}(\bk_i,z)}{\partial \theta_{\alpha}}\,\left[C_{ij}^{\rm B}\right]^{-1}\,\frac{\partial B_{\rm HI}(\bk_j,z)}{\partial \theta_{\beta}} \,,
\end{equation}
where $\theta_{\alpha}$ are the parameters of interest, $\bk_i$ is an abbreviation for the three sides of the $i$th triangle, and $\int {\rm d}^3\bk_i\equiv (4\pi)^{-1}\int_{-1}^{1}\mathrm{d}\mu_1\int_0^{2\pi}\mathrm{d}\phi\sum_T$, where $\mu_1=\hat{\bk}_1\cdot\hat{\bm z}$ and $\phi$ characterises the polar and the azimuthal orientations of a triangle relative to the line-of-sight direction, $\hat{\bm z}$. The sum $\sum_T$ runs over all triangles satisfying $k_{\rm min}\le k_3\le k_2\le k_1 \le k_{\rm max}$. The bispectrum defined in~\cref{eq:Bgs} depends on five variables: the three side lengths ($k_1$, $k_2$, $k_3$) and two for orientation relative to the line-of-sight. The forecasts presented here will use the information of the bispectrum monopole obtained after taking the average over all angles. The largest accessible scales are set by the fundamental frequency $k_{\rm f}=2\pi/L$, with $L=V_{\rm s}^{1/3}$ and $V_{\rm s}$ being the survey volume, so that $k_{\rm min}=k_{\rm f}$. For the smallest accessible scales we adopt $k_{\rm max}(z)=0.75\,k_{\rm NL}(z)$, where\footnote{A plot of $k_{\rm NL}(z)$ is shown in Ref.~\cite{Karagiannis:2022ylq}.} $k_{\rm NL}$ is defined as the inverse square root of the one-dimensional velocity dispersion~\cite{Karagiannis:2019jjx}. This confines the analysis to the perturbative regime, where the tree-level model shows good agreement with numerical results~\cite{Gil-Marin:2014sta,Lazanu2015b,Hashimoto:2017klo,Chan2017,Oddo:2019run}.
The forecast uncertainties on $\theta_\alpha$ are  given by $\sigma(\theta_\alpha)=(\tilde{\cal{I}}^{-1})_{\alpha\alpha}^{1/2}$, where $\tilde{\mathcal{I}}_{\alpha\beta} = \sum_i \mathcal{I}_{\alpha\beta} (z_i)$ encompasses the total information from all redshift bins, where they are considered independent of each other. Under the assumption of uncorrelated redshift bins, the final forecasts on the feature parameter exhibit minimal sensitivity to the choice of $\Delta z$. All the parameters that control the model are considered free and are entries of the information matrix. These include the five feature parameters, five cosmological parameters and 11 redshift-dependent nuisance parameters.\footnote{These are the HI IM bias coefficients, the AP multiplicative factors, the growth rate, the shot-noise terms and the FOG amplitudes, i.e. $\bm{\theta}(z_i)=\big\{\omega_{\rm b},\,\omega_{\rm c},\, h,\,\ln(10^{10}A_{\rm s}),\,n_{\rm s},\,{\cal A}_{\rm X},\, \omega_{\rm X},\, \phi_{\rm X}^P,\, \fnl^{\rm X},\, \phi_{\rm X}^B;\\
D_A(z_i),H(z_i),f(z_i),b_1(z_i),b_2(z_i),b_{s^2}(z_i),\sigma_P(z_i),\sigma_B(z_i),P_\veps(z_i),P_{\veps\veps_{\delta}}(z_i),B_{\veps}(z_i)\big\}.$}

For the power spectrum, we consider only the Gaussian part of the covariance
\begin{equation}
    C^{\rm P}_{kk'}=
    \frac{2(2\pi)^3}{V_{\rm s}\,V_k}\,
    P_{\rm HI}^2(k,\mu,z)\,
    \delta_{kk'} \,.
\end{equation}
For the bispectrum, the full covariance is 
\begin{equation}\label{eq:bisp_covar}
    C^{\rm B}_{ij}
    =
    C^{\rm B}_{\rm G}(\bk_i,\bk_j)
    +
    C^{\rm B}_{\rm NG}(\bk_i,\bk_j)\,,
\end{equation}
where $i,j$ label the different triangle configurations. The Gaussian contribution is given by
\begin{align} \label{eq:Cov_G}
    C_{\rm G}^{\rm B}(\bk_i,\bk_j) = \frac{(2\pi)^6}{V_{\rm s}\,V_{123}}\,s_{123}\,\delta_{ij}\,P_{\rm HI}(\bk_1,z)\,P_{\rm HI}(\bk_2,z)\,P_{\rm HI}(\bk_3,z) \,,
\end{align}
where $s_{123}=6,2,1$ for equilateral, isosceles, and scalene triangles respectively. The volume in Fourier space of the power spectra shells and the bispectra  fundamental triangle bins is, in the thin shell limit ($k\gg\Delta k$)~\cite{Sefusatti2006},
\begin{equation} \label{eq:vk}
V_k =4\pi\, k^2\,\Delta k\,, \quad 
V_{123}={8\pi^2}\,k_1\,k_2\,k_3\,\Delta k^3\,, 
\end{equation}
where the bin size is $\Delta k=k_{\rm f}$. The expression for $V_{123}$ requires some corrections for flattened ($k_1=k_2+k_3$) and open ($k_1= k_2+k_3+\Delta k$) triangles~\cite{Biagetti:2021tua}, which are also considered here. 

The non-Gaussian contribution to the bispectrum covariance is in general difficult to compute analytically, although approximate expressions exist for specific scales and triangle configurations~\cite{Barreira:2019icq,Biagetti:2021tua}. It has been shown that for certain shapes (e.g.\ squeezed triangles) and types of PNG (e.g.\ local) the effect of the non-Gaussian term can be significant~\cite{Barreira:2019icq,Biagetti:2021tua,Floss:2022wkq,Salvalaggio:2024vmx}. The feature models considered here have limited sensitivity to these configurations, as shown in~\cref{fig:SN_lin,fig:SN_log}, while the oscillatory signal in squeezed
configurations is in any case strongly damped by non-linear evolution~\cite{Goldstein:2025eyj}. Thus the non-Gaussian contribution is expected to play a lesser role. We therefore neglect ${\sf C}^B_{\rm NG}$ in our main results and assess the impact of this approximation explicitly in~\cref{sec:full_bispectrum_covariance}.

\begin{table}[t]
 \centering
 \begin{tabular}{l|c|c}
 \hline\hline
  & SKAO (Band 1) & HIRAX\\ \hline
   redshift   &  $0.35 - 3.05$ & $0.775-2.55$  \\
 $N_{\rm dish}$ & $197$ & $1,024$\\
 $D_{\rm dish}$ [m] & $15$ & $6$ \\
 $S_{\rm area}$ [$\rm{deg}^2$] & $20,000$ & $15,000$ \\
 $t_{\rm survey}$ [hrs] & $10,000$ & $17,500$ \\
  \hline\hline
 \end{tabular}
 \caption{The survey specifications for the two HI IM surveys considered in this work. }
 \label{table:survey_specs}
\end{table}

Radio telescope arrays can measure HI intensity in two ways: in single-dish (SD) mode, where each dish is auto-correlated and the resulting maps are summed, and in interferometer (IF) mode, where signals from different dishes are cross-correlated. SD surveys primarily probe large, linear scales, while IF surveys access intermediate to small scales with high angular resolution. In this work we consider SKAO\footnote{\url{www.skatelescope.org}} \cite{Bacon:2018dui} in SD mode and HIRAX\footnote{\url{hirax.ukzn.ac.za}} \cite{Crichton:2021hlc} in IF mode. The survey specifications are summarised in \cref{table:survey_specs}, where we consider for both a redshift bin of $\Delta z=0.2$. The instrumental noise is assumed Gaussian. For HIRAX it is~\cite{Zaldarriaga2003b,Tegmark2008}
\begin{equation} \label{eq:Pnoise_IF}
  P^{\rm IF}_{\rm N}(\kperp,{z})=T_{\rm sys}(z)^2\chi(z)^2\lambda(z)\frac{(1+z)}{H(z)}\left[\frac{\lambda(z)^2}{A_{\rm e}}\right]^2\frac{1}{2\,n_{\rm b}({u},z)\,t_{\rm{survey} }}\frac{S_{\rm area}}{ \theta_{\rm b}(z)^2} \,,
 \end{equation}
where $\chi$ is the comoving distance, $\lambda(z)=\lambda_{21}(1+z)$, $\theta_{\rm b}(z)=1.22\,\lambda(z)/D_{\rm dish}$ is the beam FWHM, and $A_{\rm e}=\eta\pi (D_{\rm dish}/2)^2$ with $\eta=1$. The system temperature is modelled as $T_{\rm sys}=T_{\rm inst}+T_{\rm sky}$, with $T_{\rm inst}=50\,{\rm K}$ and $T_{\rm sky}(z)/{\rm K} =25(f_{21}/[(1+z)400])^{-2.75} + 2.75$, while the baseline density $n_{\rm b}$ is taken from the simulated distributions of Ref.~\cite{Crichton:2021hlc}.

For SKAO, the SD instrumental-noise power spectrum is~\cite{Santos:2015gra}
\begin{equation} \label{eq:Pnoise_SD}
    P^{\rm SD}_{\rm N}(\kperp,z)=T_{\rm sys}(z)^2\chi(z)^2\lambda(z)\frac{(1+z)}{H(z)}\, \frac{S_{\rm area}}{\eta^2\, N_{\rm pol}\,N_{\rm dish}\,t_{\rm survey}\,{\beta_\perp}(\kperp,z)^2} \,,
\end{equation}
where we assume $\eta=1$ and $N_{\rm pol}=2$, and take $T_{\rm sys}(z)$ from Ref.~\cite{Bacon:2018dui}. The transverse beam is modeled in Fourier space as~\cite{Bull2015}
\begin{equation} \label{eq:beam}
    \beta_\perp(k,\mu,z)=\exp\left[-\frac{k_\perp^2 \chi(z)^2 \theta_{\rm b}(z)^2}{16\ln 2} \right] \!,
\end{equation}
while the radial beam can be neglected due to the high frequency resolution~\cite{Bull2015}.

Foregrounds are orders of magnitude brighter than the cosmological signal~\cite{Shaw:2013wza,Shaw:2014khi,Pober:2014lva,Byrne:2018dkh,Spinelli:2021emp}, contaminating long-wavelength radial modes~\cite{Jacobson:2003wv,Furlanetto:2006jb,Chang2007,Liu2011,Liu2012,Shaw:2013wza,Shaw:2014khi}. Although reconstruction methods can recover part of the lost information~\cite{Zhu:2016esh,Karacayli:2019iyd,Modi:2019hnu,Jasche2010,Kitaura2013,Wang:2014hia,Jasche:2014vpa,Shaw:2014khi,Wang:2016qbz,Seljak:2017rmr,Modi:2018cfi}, here we adopt a foreground-avoidance approach and exclude all scales that satisfy $\kpar <\kparmin$, taking $\kparmin=0.001\,h\,{\rm Mpc}^{-1}$ as an optimistic choice and $\kparmin=0.005\,h\,{\rm Mpc}^{-1}$ as a pessimistic one. In IF surveys, the chromatic response of the instrument leads to foreground leakage into transverse modes~\cite{Liu2011,Liu2012,Parsons2012,Pober2014,Seo2015,Pober2015}. Here we take the effect into account by excluding all modes lying in the {`foreground wedge'}, \ie requiring that
\begin{equation} \label{eq:kwedge}
    k_{\parallel}\geq A_{\rm wedge}(z)\,\kperp \,,
\end{equation}
with~\cite{Pober2014}
\begin{equation} \label{eq:wedge_prim}
    A_{\rm wedge}(z) = \frac{\chi(z)\,H(z)}{c\,(1+z)}\,\sin\!\!\left(N_{\rm w}\frac{\theta_{\rm b}(z)}{2}\right) \!,
\end{equation}
where we take $N_{\rm w}=1$. Any $k$-modes and triangle configurations outside the observational windows defined by the cuts mentioned above are excluded from the information matrix analysis. Foreground avoidance is appropriate for a forecast, but does not capture additional systematics such as polarisation leakage, radio-frequency interference, and beam effects. For recent work including such effects see Refs.~\cite{Spinelli:2021emp,Crichton:2021hlc,Fornazier:2021ini,Wang:2020lkn,Li:2020bcr,Matshawule:2020fjz,Liu:2019awk}.

\subsection{CMB forecasting methodology}
We outline the methodology employed to compute the information matrices for CMB observables. We include temperature anisotropies ($T$), $E$-mode polarisation ($E$), and the CMB lensing convergence/potential ($\phi$), and work in the Gaussian likelihood approximation for the multipole moments~\cite{Jungman:1995bz}. 

The CMB information matrix is
\begin{equation} \label{eqn:fisher_cmb}
    \mathcal{I}_{\alpha \beta}^\mathrm{CMB}
    = \sum_{\ell=\ell_\mathrm{min}}^{\ell_\mathrm{max}}
    \frac{2\ell + 1}{2}\, f_\mathrm{sky}\,
    \mathrm{Tr}\!\left[
        \frac{\partial \mathbf{C}_\ell}{\partial \theta_\alpha}\,
        \mathbf{C}_\ell^{-1}\,
        \frac{\partial \mathbf{C}_\ell}{\partial \theta_\beta}\,
        \mathbf{C}_\ell^{-1}
    \right] \!,
\end{equation}
where $\{\theta_\alpha\}$ denotes the set of cosmological parameters, $f_\mathrm{sky}$ is the observed sky fraction, and the trace is over the $\{T,E,\phi\}$ basis. The theoretical covariance matrix is
\begin{equation} \label{eq:cmb_cov}
    \mathbf{C}_\ell =
    \begin{pmatrix}
        C_\ell^{TT} + \mathcal{N}_\ell^{TT} & C_\ell^{TE} & C_\ell^{T\phi} \\
        C_\ell^{TE} & C_\ell^{EE} + \mathcal{N}_\ell^{EE} & C_\ell^{E\phi} \\
        C_\ell^{T\phi} & C_\ell^{E\phi} & C_\ell^{\phi\phi} + \mathcal{N}_\ell^{\phi\phi}
    \end{pmatrix} \!,
\end{equation}
where $\mathcal{N}_\ell^{XX}$ denotes the instrumental noise (including beam effects) and $\mathcal{N}_\ell^{\phi\phi}$ is the reconstruction noise for lensing (we set noise cross-correlations to zero and include signal cross-spectra such as $C_\ell^{T\phi}$ and $C_\ell^{E\phi}$).

Following the standard approximation of Ref.~\cite{Knox:1995dq}, the temperature and polarisation noise spectra are
\begin{equation}
    \mathcal{N}_{\ell}^{XX}
    = w_{XX}^{-1}\,
    \exp\!\left[
        \frac{\ell(\ell+1)\,\theta_\mathrm{FWHM}^2}{8 \ln 2}
    \right] \!,
\end{equation}
with $X\in\{T,E\}$. Here $\theta_\mathrm{FWHM}$ is the beam full-width at half-maximum (in radians), while $w_{TT}$ and $w_{EE}$ are the inverse noise variances (typically quoted in $(\mu\mathrm{K}\,\mathrm{arcmin})^{-2}$).
For the lensing noise term $\mathcal{N}_\ell^{\phi\phi}$, we assume reconstruction with a quadratic estimator built from
the observed temperature and polarisation maps. We compute $\mathcal{N}_\ell^{\phi\phi}$ using the minimum-variance estimator
formalism of Ref.~\cite{Okamoto:2003zw} as implemented in the \texttt{quicklens} package.\footnote{\url{https://github.com/dhanson/quicklens}}

We model a next-generation ground-based experiment using the Simons Observatory (SO) noise curves \citep{SimonsObservatory:2018koc},
including the impact of component-separation residuals.
We adopt $f_\mathrm{sky}=0.4$ for SO and use multipoles $\ell\in[40,3000]$ for $TT$ and $TE$, and $\ell\in[40,5000]$ for $EE$.
For CMB lensing we include $\ell\in[2,3000]$.
Since ground-based experiments are limited on the largest angular scales, we complement the SO information with LiteBIRD~\cite{LiteBIRD:2022cnt} for $\ell\in[2,40]$ over a sky fraction $f_\mathrm{sky}=0.7$ \cite{Bermejo-Climent:2021jxf}.
In practice, we compute separate information matrices for SO and LiteBIRD on their respective multipole ranges and
add them, consistent with the independence approximation adopted in~\cref{eqn:infomatrix_sum}.

In this analysis we treat CMB measurements as statistically independent from the other probes considered; therefore the total information is obtained by direct summation,
\begin{equation} \label{eqn:infomatrix_sum}
    \mathcal{I}^\mathrm{tot} = \mathcal{I}^\mathrm{other} + \mathcal{I}^\mathrm{CMB} \,.
\end{equation}
Here $\mathcal{I}^\mathrm{other}$ includes the information matrix for both the HI IM power spectrum and bispectrum, \cref{eq:fisherPs,eq:fisherBs}.
In the CMB block we include three feature parameters (i.e.\ those affecting the primordial power spectrum) and six cosmological parameters.
We marginalise over the reionization optical depth $\tau_\mathrm{reio}$ in the CMB information matrix before combining it with the HI IM information matrices.

\section{Results} \label{sec:results}
This section presents forecasted uncertainties on primordial feature models from the SKAO and HIRAX HI IM surveys, together with forecasts from upcoming CMB experiments. For HI IM, results are obtained from the power spectrum, the bispectrum, and their combination, considering both optimistic and pessimistic foreground scenarios. For the CMB, we present angular power spectrum forecasts from the combined sensitivity of SO and LiteBIRD. Results for the linear and logarithmic templates are shown in~\cref{fig:main_results_lin_log} and summarised in~\cref{table:lin_log_results}, while for the sharp-feature model results are shown in~\cref{fig:main_results_gsr} and summarised in~\cref{table:gsr_results}. 

In the combined power spectrum and bispectrum ($P+B$) analysis, the cross-covariance is neglected, as it is expected to have a minimal impact on the final sensitivity, see Refs.~\cite{Chan2017,Yankelevich:2018uaz,Barreira:2019icq,Biagetti:2021tua}. We also include IR resummation in the modelling of both HI IM correlators (see~\Cref{sec:HI_model}), which for the scales and redshifts considered here, has a minimal impact on the forecasted constraints.

\subsection{Linear and logarithmic features}
\begin{figure}[t]
    \centering
    \includegraphics[width=\textwidth]{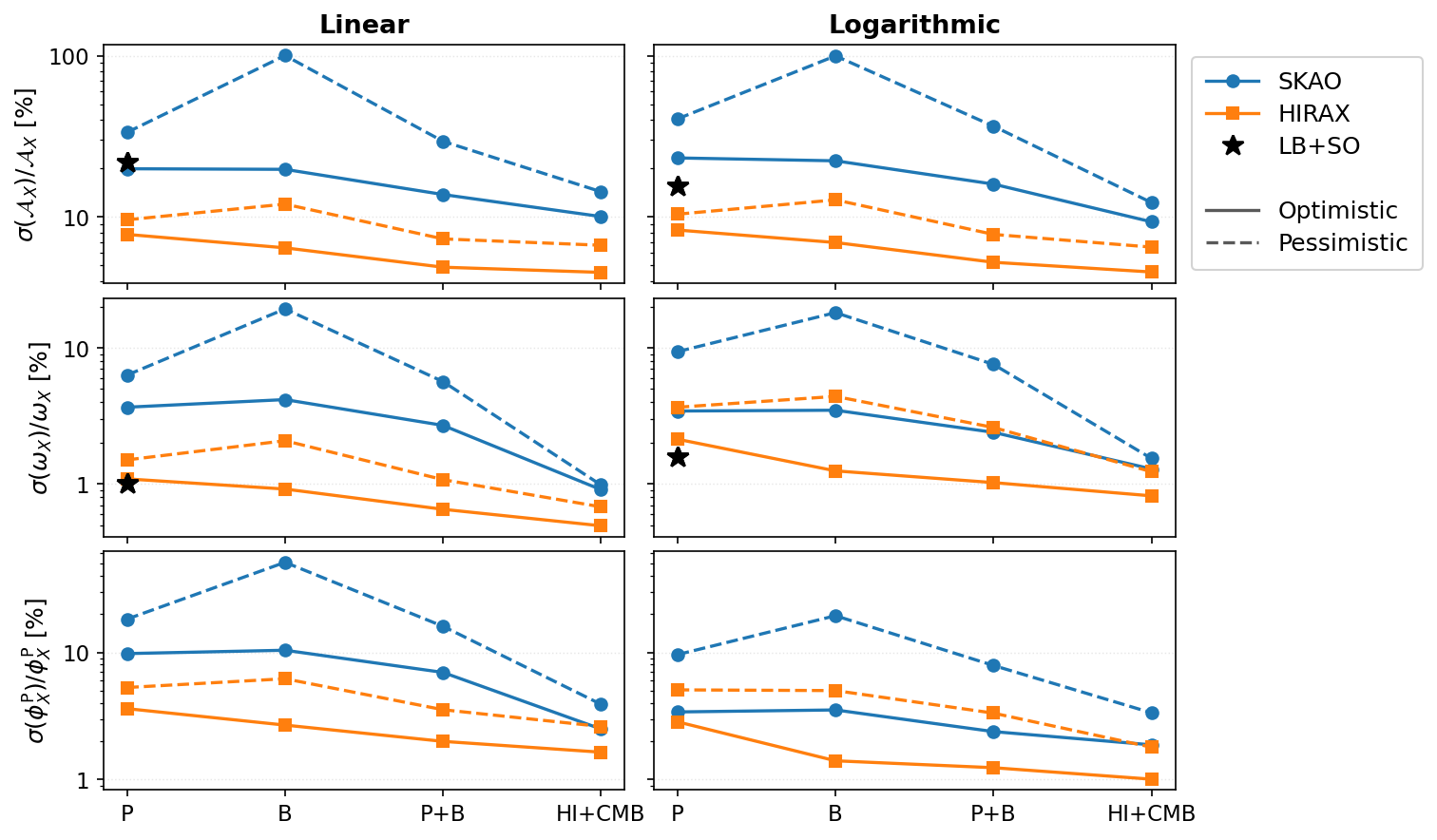}
    \caption{Forecasted 68\% CL marginalised relative uncertainties
on the power spectrum feature parameters ${\cal A}_X$ (amplitude, top row),
$\omega_X$ (frequency, middle row), and $\phi_X^{\rm P}$ (phase, bottom row)
from SKAO (circles), HIRAX (squares), and SO+LiteBIRD (stars),
using the power spectrum ($P$), bispectrum ($B$), their combination ($P+B$),
and the joint 21\,cm and CMB analysis (HI+CMB).
The left and right columns correspond to the linear and logarithmic oscillatory
models, respectively.
Solid and dashed lines correspond to the optimistic
($k_{\parallel,\rm min} = 0.001\,h\,{\rm Mpc}^{-1}$) and pessimistic
($k_{\parallel,\rm min} = 0.005\,h\,{\rm Mpc}^{-1}$) foreground cuts,
respectively. }\label{fig:main_results_lin_log}
\end{figure}
For the linear and logarithmic feature templates, HIRAX generally outperforms SKAO across most of the parameter space, as shown in \cref{fig:main_results_lin_log}. In the optimistic foreground case, the HIRAX power spectrum constraint on the feature amplitude is approximately 60\% tighter than the corresponding SKAO constraint for both templates. In the combined $P+B$ analysis, HIRAX improves over SKAO by roughly $65\%$ for $\mathcal{A}_\mathrm{lin}$ and by about $70\%$ for $\mathcal{A}_\mathrm{log}$. 

The performance of HIRAX is expected from its interferometric configuration, which effectively probes transverse intermediate and small scales. At these scales, oscillatory signals benefit from a higher density of accessible modes and triangle configurations. Conversely, SKAO, operating in single-dish mode, is more susceptible to angular resolution limits and the loss of long radial modes due to foreground cuts. This is evident when moving to pessimistic foreground assumptions: the SKAO $P+B$ constraint on linear feature amplitude degrades by a factor of two, whereas the HIRAX constraint degrades by $\sim 40\%$. For the logarithmic feature amplitude, SKAO degrades by more than a factor of two, while HIRAX degrades by $\sim 60\%$. This behaviour is consistent with previous PNG forecasts, where HI bispectrum uncertainties were found to benefit particularly from interferometric surveys, see Refs.~\cite{Karagiannis:2019jjx,Karagiannis:2020dpq,Karagiannis:2024pyx}.

The bispectrum contributes to the uncertainties in two ways. Firstly, it is the only observable considered here that directly constrains the non-Gaussian feature parameters $f_{\rm NL}^{X}$ and $\phi_X^{\rm B}$. For both templates, HIRAX improves on SKAO by about 55--60\% for these parameters in the optimistic case, despite the fact that the fiducial $f_{\rm NL}^{X}=1$ is not detected at high significance in this setup. Secondly, the bispectrum improves  constraints on parameters present in the power spectrum, as previously shown and discussed in Ref.~\cite{Euclid:2023shr}. This occurs because the late-time gravitational contribution to the tree-level bispectrum depends on products of the power spectra, see~\cref{eq:Bgs}, allowing the bispectrum to retain sensitivity to the oscillatory features even in the absence of a primordial bispectrum contribution (see the right column of~\cref{fig:fid_templates}). The $P+B$ combination tightens the amplitude uncertainties by about $30$--$40\%$. The gain is also clear for the phase: for HIRAX, the constraint improves by roughly a factor of two, while SKAO shows a smaller but still noticeable improvement. This indicates that the improvement in $\phi_X^{\rm P}$ is driven mainly by bispectrum information, especially for HIRAX, rather than by the power spectrum alone. By contrast, the oscillation frequency is already well constrained by the power spectrum alone, so the additional gain from the bispectrum is modest, especially for HIRAX where percent-level precision is already reached.

\begin{table*}[t]
\centering
\resizebox{\textwidth}{!}{
\begin{tabular}{l ccc ccc c}
\hline\hline
 & \multicolumn{3}{c}{\bf SKAO} 
 & \multicolumn{3}{c}{\bf HIRAX}
 & {\bf CMB} \\
Parameter & $P$ & $B$ & $P+B$ & $P$ & $B$ & $P+B$ & $P$ \\
\hline
\\[-0.8em]
\multicolumn{8}{c}{\bf Linear} \\[0.3em]
\hline
$\sigma({\cal A}_{\rm lin})/{\cal A}_{\rm lin}$ 
& 0.20 (0.33) & 0.20 (1.0) & 0.14 (0.29) 
& 0.08 (0.10) & 0.06 (0.12) & 0.05 (0.07)
& 0.22 \\

$\sigma(\omega_{\rm lin})/\omega_{\rm lin}$ 
& 0.04 (0.06) & 0.04 (0.19) & 0.03 (0.06) 
& 0.01 (0.02) & 0.01 (0.02) & 0.01 (0.01)
& 0.01 \\

$\sigma(\phi_{\rm lin}^{\rm P})$ 
& 0.10 (0.18) & 0.10 (0.51) & 0.07 (0.16) 
& 0.04 (0.05) & 0.03 (0.06) & 0.02 (0.04)
& -- \\

$\sigma(f_{\rm NL}^{\rm lin})/f_{\rm NL}^{\rm lin}$ 
& -- & 19.0 (73.8) & 19.0 (73.6) 
& -- & 7.9 (15.8) & 7.9 (15.8)
& -- \\

$\sigma(\phi_{\rm lin}^{\rm B})$ 
& -- & 3.8 (15.0) & 3.8 (14.9) 
& -- & 1.6 (3.2) & 1.6 (3.2)
& -- \\[0.3em]
\hline
\\[-0.8em]
\multicolumn{8}{c}{\bf Logarithmic} \\[0.3em]
\hline
$\sigma({\cal A}_{\rm log})/{\cal A}_{\rm log}$ 
& 0.23 (0.41) & 0.22 (1.0) & 0.16 (0.37)
& 0.08 (0.10) & 0.07 (0.13) & 0.05 (0.08)
& 0.15 \\

$\sigma(\omega_{\rm log})/\omega_{\rm log}$ 
& 0.03 (0.09) & 0.03 (0.18) & 0.02 (0.08)
& 0.02 (0.04) & 0.01 (0.04) & 0.01 (0.03)
& 0.02 \\

$\sigma(\phi_{\rm log}^{\rm P})$ 
& 0.03 (0.10) & 0.04 (0.19) & 0.02 (0.08)
& 0.03 (0.05) & 0.01 (0.05) & 0.01 (0.03)
& -- \\

$\sigma(f_{\rm NL}^{\rm log})/f_{\rm NL}^{\rm log}$ 
& -- & 19.9 (131.5) & 19.7 (112.3)
& -- & 8.7 (20.3) & 8.6 (19.6)
& -- \\

$\sigma(\phi_{\rm log}^{\rm B})$ 
& -- & 4.0 (26.7) & 4.0 (20.4)
& -- & 1.7 (4.1) & 1.7 (4.0)
& -- \\[0.3em]
\hline\hline
\end{tabular}
}
\caption{Forecasted 68\% CL uncertainties for the linear and logarithmic templates from the power spectrum ($P$), bispectrum ($B$), and their combination ($P+B$). Values in parentheses correspond to the pessimistic foreground cut. The CMB column lists angular power spectrum uncertainties from SO+LiteBIRD.}
\label{table:lin_log_results}
\end{table*}

The CMB remains a competitive benchmark for the oscillation frequency, reaching precision comparable to HI IM. However, for the feature amplitude, SO+LiteBIRD is generally outperformed by the HIRAX $P+B$ analysis. In the linear template, HIRAX improves upon the CMB amplitude constraint by roughly 75\%. The CMB power spectrum provides negligible uncertainties on the phase $\phi_X$, highlighting the importance of LSS data. The joint HI+CMB analysis is most impactful for SKAO under pessimistic foregrounds, while for HIRAX the relative gain is smaller as its internal uncertainties are already dominant. Nevertheless, the joint analysis HI+CMB remains the most robust option, as the probes cover different scales, redshifts, and systematics.

As a simple robustness test, we also repeat the forecasts for a feature amplitude ten times larger than the fiducial value ${\cal A}_X = 0.1$. The corresponding constraints are reported in~\cref{app:Amplx10}, in~\cref{tab:lin_log_results_Amplx10} and \cref{tab:gsr_results_Amplx10} for the linear/logarithmic and sharp-feature cases, respectively. As expected, the feature parameters are constrained more tightly in this case, with the CMB constraints on the feature amplitude and frequency becoming substantially stronger than in the fiducial analysis. The qualitative picture, however, is unchanged: increasing the feature amplitude mainly rescales the overall constraining power, without altering the relative ranking of the surveys or the complementarity between the power spectrum, bispectrum, and CMB contributions.

\begin{figure}[t]
    \centering
    \includegraphics[width=\textwidth]{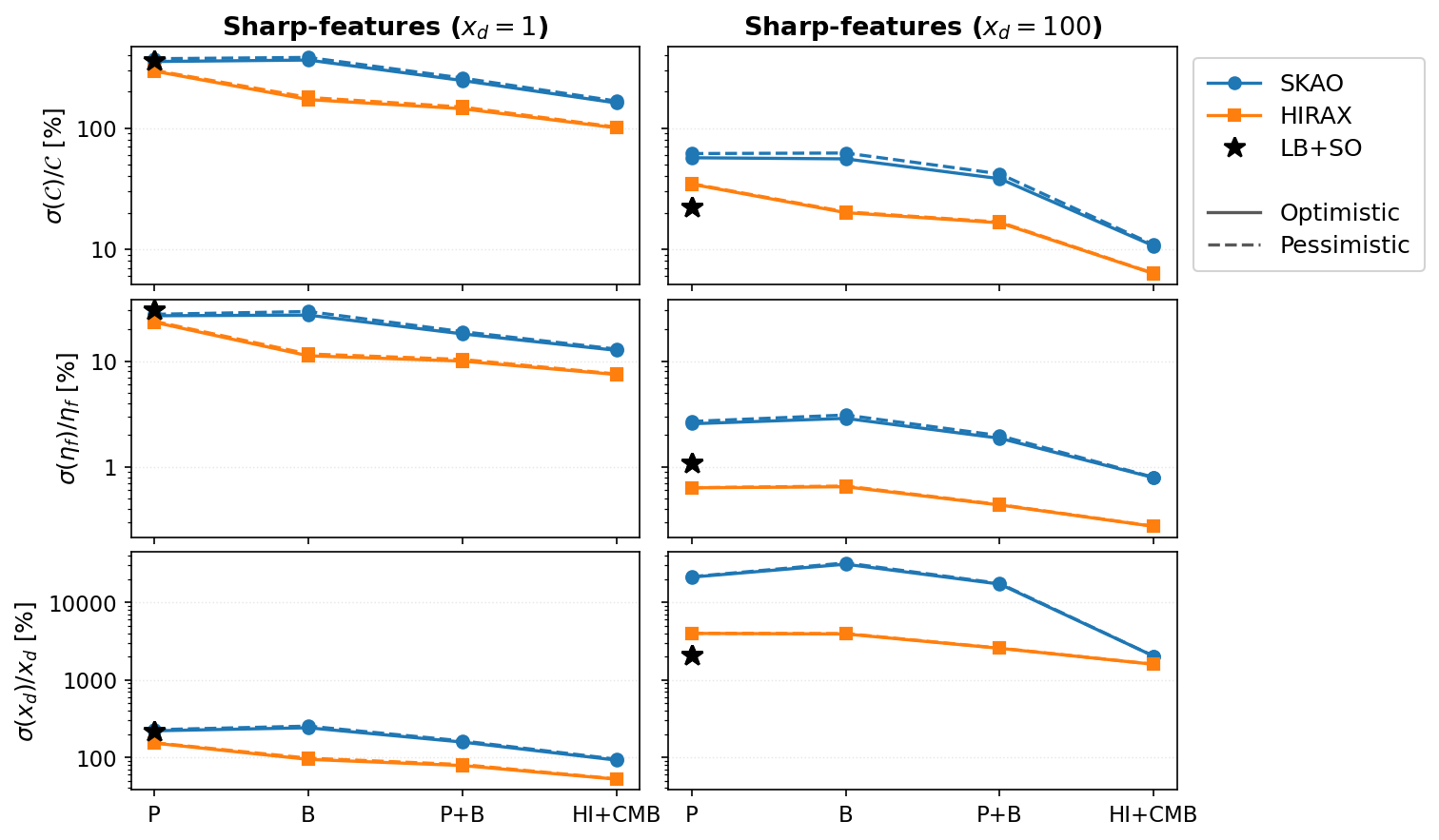}
\caption{As in~\cref{fig:main_results_lin_log}, but for the sharp-feature model parameters. The left and right columns correspond to $x_d = 1$ (sharp transition) and $x_d = 100$ (broad transition), respectively. }
    \label{fig:main_results_gsr}
\end{figure}

\subsection{Sharp-feature model}
The step model provides a consistent description of features where the power spectrum and bispectrum are linked to a single departure from slow roll connected to a transition in the inflationary potential. Unlike the linear and logarithmic templates, the power spectrum and bispectrum signals are tied to the same inflationary dynamics. The model is described by the feature amplitude $\mathcal{C}$, the conformal time of the feature $\eta_f$, and the width parameter $x_d$, which controls the sharpness of the transition. We consider two fiducial cases, $x_d^{\rm fid}=1$ (sharp transition) and $x_d^{\rm fid}=100$ (broad transition). %

For $x_d^{\rm fid}=1$, the uncertainties are generally weaker than for the previous templates, but the same survey hierarchy remains. HIRAX outperforms SKAO for all three feature parameters, especially once the bispectrum is included. In the optimistic foreground case, the HIRAX $P+B$ uncertainties are roughly $40$--$50\%$ tighter than SKAO, depending on the parameter. The bispectrum is particularly useful for HIRAX, improving the uncertainties by about $50\%$ for $\mathcal{C}$ and $x_d$, and by nearly $60\%$ for $\eta_f$, compared to a more moderate $\sim 30\%$ gain for SKAO. This again reflects the fact that the interferometric survey benefits more from the large number of small- and intermediate-scale triangle configurations, while the single-dish setup is more limited by the loss of radial modes and angular resolution. For $x_d^{\rm fid}=100$, the uncertainties on $\mathcal{C}$ and $\eta_f$ improve substantially, indicating that this configuration produces a more interesting target. HIRAX again gives the strongest HI uncertainties: in the $P+B$ combination it improves over SKAO by more than $50\%$ for $\mathcal{C}$ and more than $70\%$ for $\eta_f$. The bispectrum remains most useful for the amplitude, improving the constraint by about $30\%$ for SKAO and about $50\%$ for HIRAX. The gain for $\eta_f$ is smaller, since the power spectrum already captures much of the information on the oscillation scale. On the other hand, $x_d$ remains poorly constrained for $x_d^{\rm fid}=100$ even after combining $P+B$. The bispectrum still helps, reducing the uncertainty by about $20\%$ for SKAO and $35\%$ for HIRAX, but the final uncertainties remain weak, suggesting a strong degeneracy between the width of the feature and the parameters controlling the oscillatory envelope.

Comparison with the CMB for the sharp-feature model, see~\cref{table:gsr_results}, shows that for $x_d^{\rm fid}=1$, HIRAX $P+B$ outperforms SO+LiteBIRD by $60$--$70\%$. For $x_d^{\rm fid}=100$, the CMB becomes more competitive, particularly for $x_d$, where it slightly exceeds HIRAX sensitivity. The joint HI+CMB analysis significantly improves SKAO results, with $x_d$ uncertainties tightening by nearly an order of magnitude.

\begin{table*}[t]
\centering
\resizebox{\textwidth}{!}{
\begin{tabular}{l ccc ccc c}
\hline\hline
 & \multicolumn{3}{c}{\bf SKAO} 
 & \multicolumn{3}{c}{\bf HIRAX}
 & {\bf CMB} \\
Parameter & $P$ & $B$ & $P+B$ & $P$ & $B$ & $P+B$ & $P$ \\
\hline
\\[-0.8em]
\multicolumn{8}{c}{$\mathbf{Sharp\ feature},\, x_d^{\rm fid}=1$} \\[0.3em]
\hline
$\sigma(C)/C$ 
& 3.5 (3.7) & 3.6 (3.8) & 2.5 (2.6)
& 3.0 (3.0) & 1.7 (1.8) & 1.4 (1.5)
& 3.6 \\

$\sigma(\eta_f)/\eta_f$ 
& 0.27 (0.28) & 0.27 (0.29) & 0.18 (0.19)
& 0.23 (0.24) & 0.11 (0.12) & 0.10 (0.10)
& 0.30 \\

$\sigma(x_d)/x_d$ 
& 2.2 (2.3) & 2.4 (2.5) & 1.6 (1.6)
& 1.5 (1.5) & 0.94 (0.97) & 0.78 (0.81)
& 2.1 \\[0.3em]
\hline
\\[-0.8em]
\multicolumn{8}{c}{$\mathbf{Sharp\ feature},\, x_d^{\rm fid}=100$} \\[0.3em]
\hline
$\sigma(C)/C$ 
& 0.57 (0.62) & 0.56 (0.62) & 0.38 (0.42)
& 0.34 (0.35) & 0.20 (0.20) & 0.16 (0.17)
& 0.22 \\

$\sigma(\eta_f)/\eta_f$ 
& 0.026 (0.027) & 0.029 (0.031) & 0.019 (0.020)
& 0.006 (0.006) & 0.006 (0.007) & 0.004 (0.004)
& 0.011 \\

$\sigma(x_d)/x_d$ 
& 212.3 (214.9) & 311.7 (323.9) & 172.5 (175.9)
& 39.8 (40.0) & 39.2 (39.6) & 25.6 (25.7)
& 20.5 \\[0.3em]
\hline\hline
\end{tabular}
}
\caption{As in \Cref{table:lin_log_results}, but for the sharp-feature model.}
\label{table:gsr_results}
\end{table*}

\subsection{Frequency and scale dependence}

\begin{figure}[t]
    \centering
    \includegraphics[width=\textwidth]{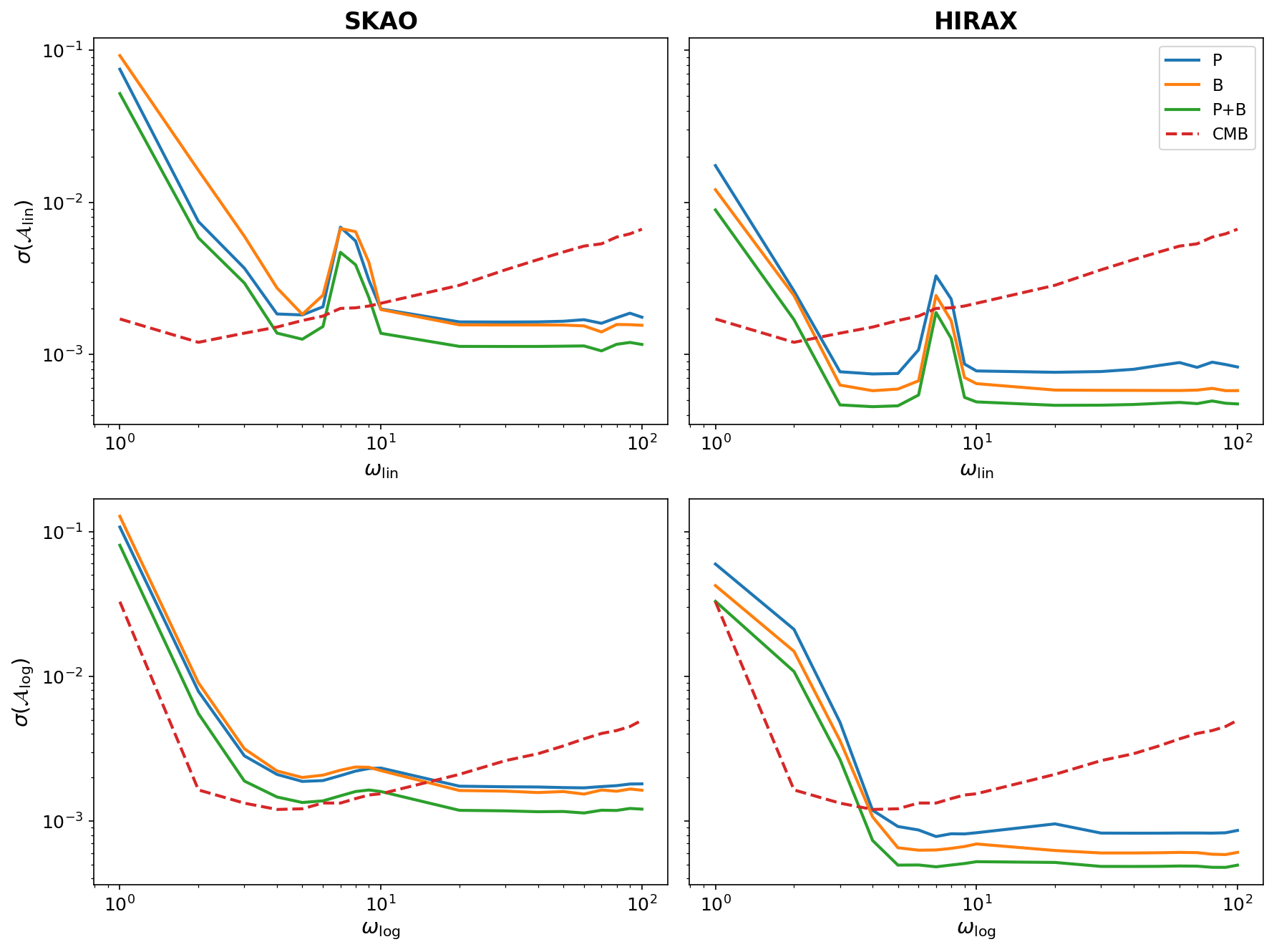}
    \caption{Marginalised 68\% CL uncertainties on the primordial feature amplitude $\mathcal{A}_X$ as a function of the frequency $\omega_X$. The upper panels refer to the linear model ($X=\mathrm{lin}$), while the lower panels refer to the logarithmic model ($X=\mathrm{log}$). We display uncertainties derived from the two HI IM surveys considered, as well as from the CMB power spectrum (red dashed line).}
    \label{fig:sigma_Ap(omega)}
\end{figure}

\begin{figure}[t]
    \centering
    \includegraphics[width=\textwidth]{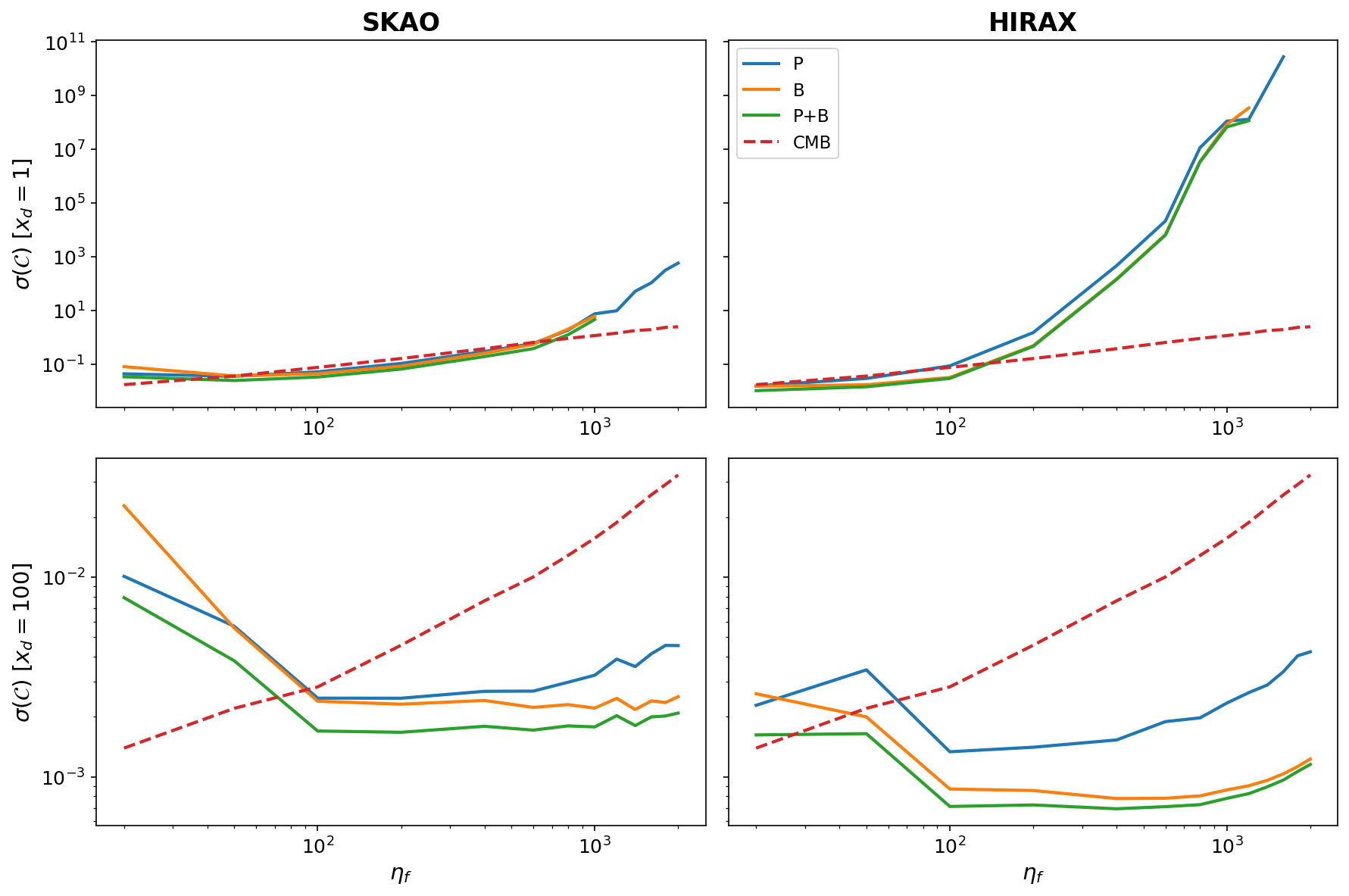}
\caption{As in \Cref{fig:sigma_Ap(omega)}, but for the sharp-feature model. The upper and lower panels correspond to $x_d = 1$ (sharp transition) and $x_d = 100$ (broad transition), respectively, with the uncertainties on the amplitude $\mathcal{C}$ plotted as a function of the oscillation frequency $\eta_f$.}
        \label{fig:sigma_C(etaf)}
\end{figure}

The dependence of amplitude uncertainties on the frequency value is shown in~\cref{fig:sigma_Ap(omega),fig:sigma_C(etaf)}. For the linear template, uncertainties improve until $\omega_{\rm lin}\sim 4$--$6$. Beyond this, uncertainties enter a plateau for $\omega_{\rm lin}\gtrsim 10$, indicating information saturation. A localised degradation occurs around $\omega_{\rm lin}\sim 7$--$9$, where the primordial frequency overlaps with the BAO scale; see Refs.~\cite{Beutler:2019ojk,Ballardini:2019tuc} for previous discussions.
The logarithmic template exhibits a smoother trend, with a less pronounced BAO-induced degradation.
For the sharp-feature model, uncertainties for $x_d^{\rm fid}=1$ degrade rapidly at large $\eta_f$, as high-frequency oscillations are not resolved by the survey $k$-binning. This effect is more severe for the bispectrum, where the oscillatory signal cancels more efficiently across triangle configurations, potentially leading to a singular information matrix in high-redshift bins. Conversely, for $x_d^{\rm fid}=100$, uncertainties plateau for $\eta_f \gtrsim 100$, indicating that the feature is well-contained within the survey's sensitive range.

The origin of this frequency dependence can be seen more directly in the triangle-space S/N maps shown in Appendix~\ref{app:sn_triangles}. As $\omega_{\rm lin}$ increases, the oscillatory S/N bands move across the triangle domain. For SKAO this mainly redistributes the signal among accessible configurations, whereas for HIRAX the foreground wedge and interferometric window remove part of the domain, making the measured S/N more sensitive to whether the high-S/N bands fall inside the accessible region. This explains the non-monotonic behaviour of the amplitude constraints with frequency. These appendix plots are shown only for the linear template, as an illustrative test of how the oscillatory pattern changes with frequency and how this affects its overlap with the survey window; the same qualitative interpretation applies to the other feature models.

In general, CMB data outperform radio surveys for low frequencies. This is crucial since the perturbative description of the non-linear model breaks down for broad oscillations; see Refs.~\cite{Beutler:2019ojk,Ballardini:2019tuc,Ballardini:2024dto}.

\begin{figure*}[t]
    \centering

{\large\bfseries SKAO}\\
\includegraphics[width=\textwidth]{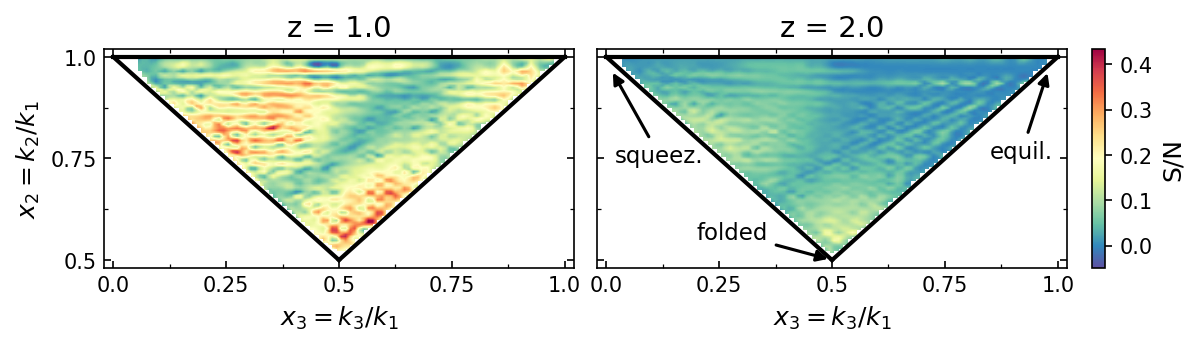}

{\large\bfseries HIRAX}\\
\includegraphics[width=\textwidth]{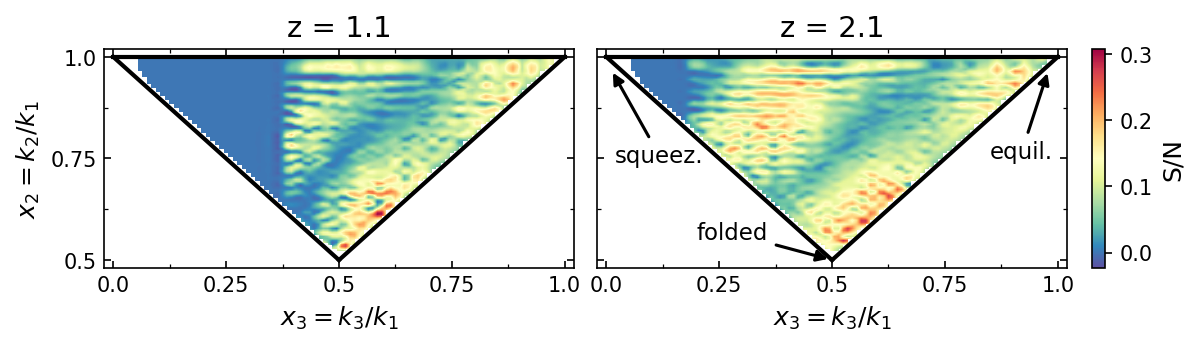}

    \caption{Angular-averaged signal-to-noise contribution of bispectrum triangle configurations for the linear feature model. The upper and lower panels correspond to SKAO and HIRAX, respectively. Triangle shapes are described by $x_2 = k_2/k_1$ and $x_3 = k_3/k_1$, with $k_1 \geq k_2 \geq k_3$. The edges and vertices of the allowed region highlight the squeezed, folded, and equilateral limits. The colour scale shows the SNR associated with each triangle shape after applying the survey window and foreground cuts.}
    \label{fig:SN_lin}
\end{figure*}

\subsection{Bispectrum signal-to-noise ratio and triangle configurations}
\begin{figure*}[t]
    \centering
    {\large\bfseries SKAO}\\    
    \includegraphics[width=\textwidth]{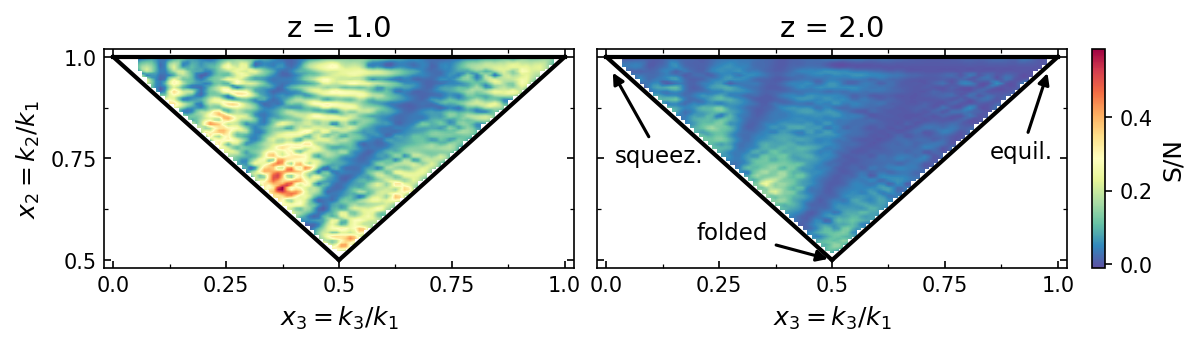}

    {\large\bfseries HIRAX}\\
    \includegraphics[width=\textwidth]{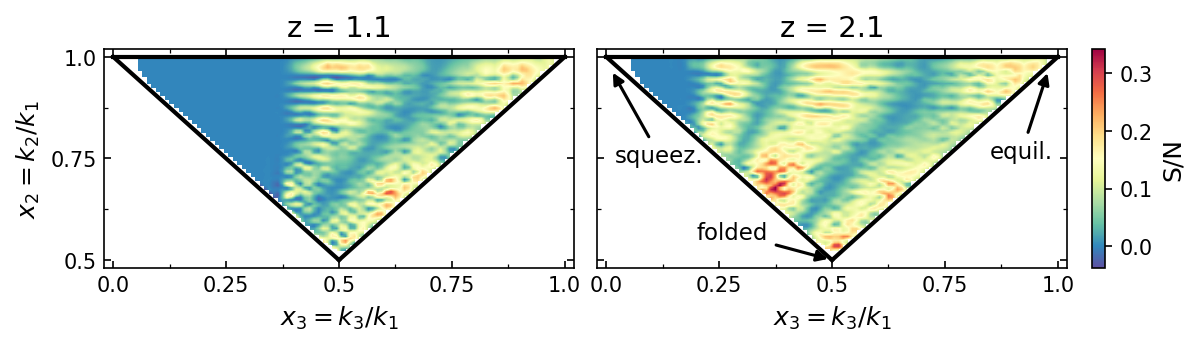}

    \caption{
    As in \Cref{fig:SN_lin}, but for the logarithmic feature model.
    }
    \label{fig:SN_log}
\end{figure*}

\begin{figure*}[t]
    \centering
    {\large\bfseries SKAO}\\ 
    \includegraphics[width=\textwidth]{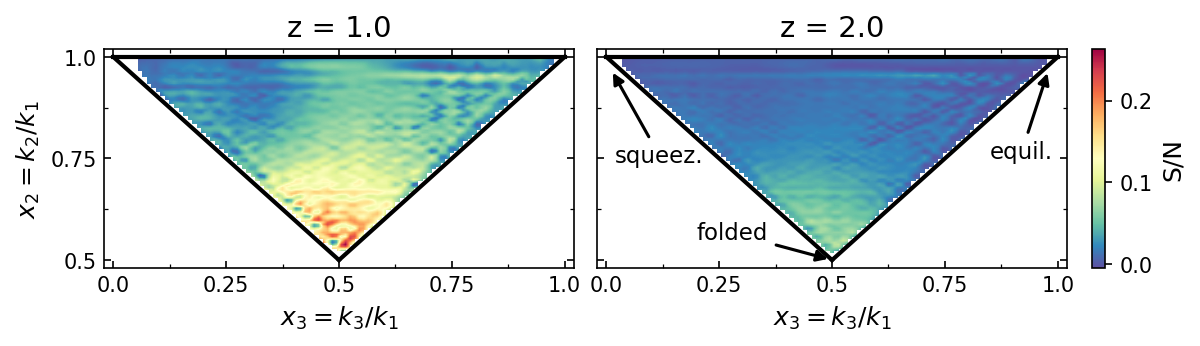}

    {\large\bfseries HIRAX}\\
    \includegraphics[width=\textwidth]{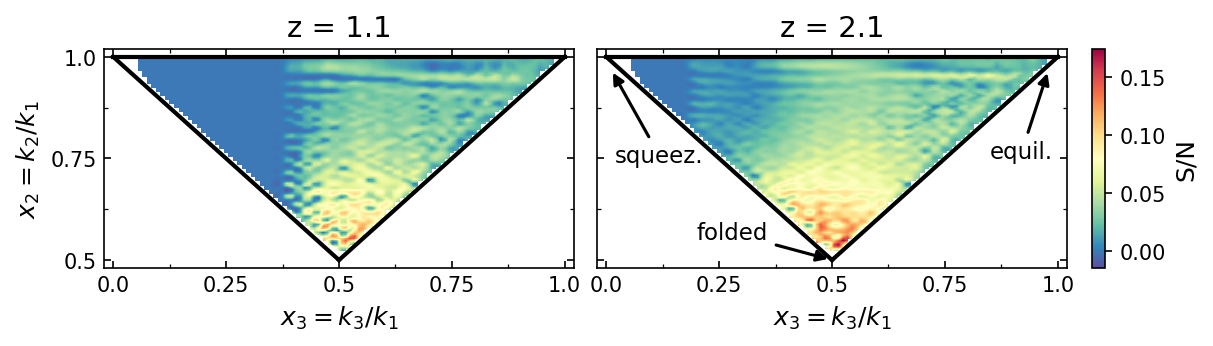}

    \caption{
    As in \Cref{fig:SN_lin}, but for the sharp-feature model with $x_d=1$.
    }
    \label{fig:SN_gsr_xd_1}
\end{figure*}

\begin{figure*}[t]
    \centering
    {\large\bfseries SKAO}\\ 
    \includegraphics[width=\textwidth]{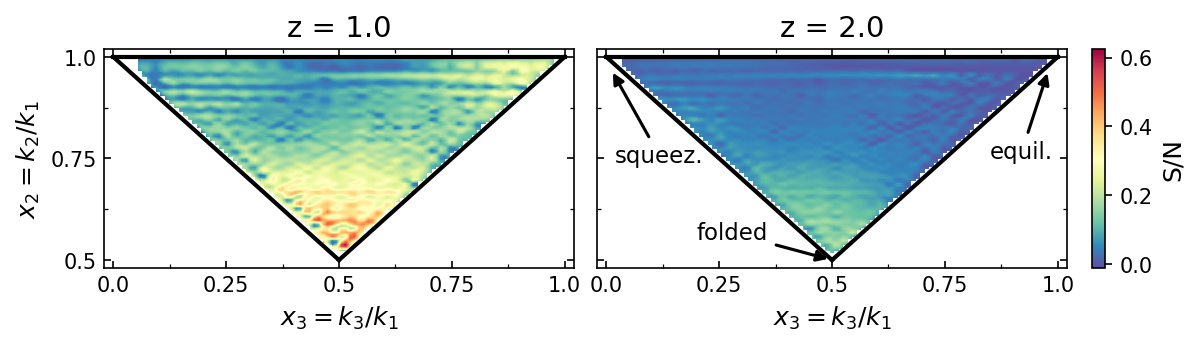}

    {\large\bfseries HIRAX}\\
    \includegraphics[width=\textwidth]{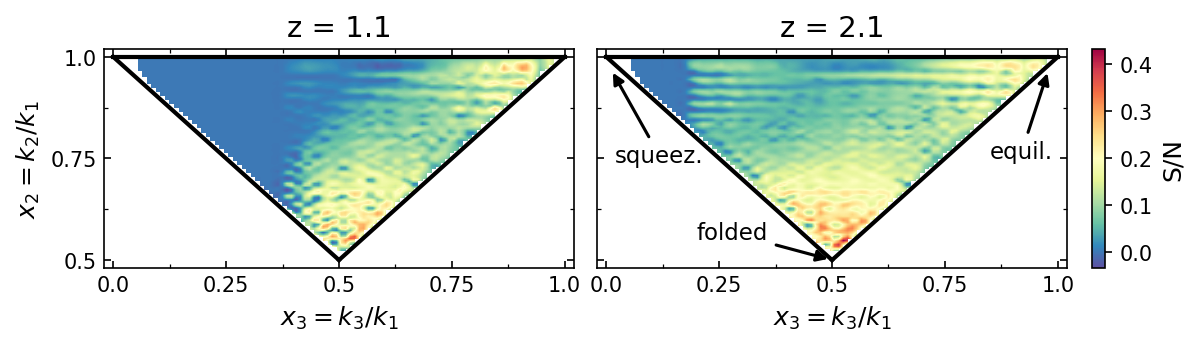}

    \caption{
    As in \Cref{fig:SN_lin}, but for the sharp-feature model with $x_d=100$.
    }
    \label{fig:SN_gsr_xd_100}
\end{figure*}

Given the importance of the bispectrum for constraining feature models with LSS surveys, we examine its dependence on triangle configurations. To quantify this, the feature bispectrum SNR is computed as
\begin{equation}\label{eq:snrng}
({\rm S/N})_{\rm feat}^2(z)
=
\Big\langle
\Delta B_{\rm HI}^{\rm feat}(\mathbf{k}_i,z)\,
\left[C^{\rm B}_{ij}(z)\right]^{-1}\,
\Delta B_{\rm HI}^{\rm feat}(\mathbf{k}_j,z)
\Big\rangle_{\Omega}\,,
\end{equation}
where
\[
\Delta B_{\rm HI}^{\rm feat}(\mathbf{k},z)
\equiv
B_{\rm HI}^{\rm obs}(\mathbf{k},z)\big|_{\rm feat}
-
B_{\rm HI}^{\rm obs}(\mathbf{k},z)\big|_{\rm G}
\]
is the component of the observed HI bispectrum sourced by the primordial feature, defined as the difference between the feature model calculated at the fiducial parameters and the Gaussian baseline. The notation $\langle \cdots \rangle_{\Omega}$ denotes the average over triangle orientations with respect to the line of sight.
The SNR distribution across triangle shapes is shown in~\cref{fig:SN_lin,fig:SN_log,fig:SN_gsr_xd_1,fig:SN_gsr_xd_100}, parametrised by $x_2=k_2/k_1$ and $x_3=k_3/k_1$ with $k_1\geq k_2\geq k_3$. Squeezed configurations are located near the upper-left corner, equilateral triangles near the upper-right corner, folded configurations along the lower boundary, and elongated or isosceles shapes along the sides of the domain. The colour scale indicates the angular-averaged SNR after applying the survey window and foreground cuts.

The signal is non-uniform across triangle space, manifesting as oscillatory bands whose position and contrast depend on the feature template, redshift, and survey window. Varying the triangle shape alters both the primordial bispectrum contribution and the late-time gravitational terms, meaning different configurations sample different phases of the oscillatory signal. The bispectrum constraining power is consequently determined by the feature amplitude, the total number of accessible triangles, and the overlap between the oscillatory pattern and the survey window.

For the linear template (\Cref{fig:SN_lin}), the SNR shows that the bispectrum signal extends over a broad range of triangle shapes, displaying oscillatory structures from folded to equilateral configurations. The largest contributions arise primarily from folded and moderately flattened triangles, with a non-zero contribution near the equilateral limit. The signal is not dominated by squeezed configurations; instead, the primary information originates from elongated and isosceles shapes, where the oscillatory dependence on $K$ is sampled across a large number of available configurations. This accounts for the performance of HIRAX, as its interferometric configuration provides access to a large number of transverse modes, and thus to the triangle configurations carrying the largest feature signal.

For SKAO, the SNR pattern varies between the two redshift bins. At lower redshift, the signal is distributed across a wider domain, whereas at higher redshift the accessible region is constrained by the survey window and foreground cuts. For HIRAX at low redshift, a significant part of the triangle domain is removed by the interferometric window and foreground wedge (represented by the blank region in the plots). The remaining region contains an oscillatory signal towards folded and equilateral-like configurations. At higher redshift, the HIRAX window shifts to a different portion of triangle space, altering the signal distribution. The optimal triangle configurations are therefore survey- and redshift-dependent, implying that a detection strategy based on a fixed subset of triangle shapes is sub-optimal.

The logarithmic template (\Cref{fig:SN_log}) exhibits a similar qualitative behaviour, but with a modified pattern of oscillatory bands. Compared to the linear case, the signal varies less rapidly for a fixed change in $K$ due to the scale-dependent period of logarithmic oscillations. Consequently, the high-SNR regions are redistributed across triangle space. For SKAO, the dominant contributions come from folded and elongated triangles, with squeezed configurations being subdominant. For HIRAX, the signal is concentrated within the region permitted by the interferometric window, with contributions spanning from elongated to equilateral shapes. This indicates that the bispectrum information for feature models is distributed across an extended set of shapes whose phases align with the oscillatory signal, rather than being concentrated in a single limiting configuration.

The sharp-feature model results in \Cref{fig:SN_gsr_xd_1,fig:SN_gsr_xd_100} exhibit a more scale-localised oscillatory signal because the power spectrum and bispectrum are generated by the same inflationary transition. For $x_d^{\rm fid}=1$, the SNR is concentrated primarily around folded and moderately flattened triangles, particularly at lower redshifts. The overall signal amplitude is lower than in the linear and logarithmic templates, corresponding to the larger marginalised uncertainties reported in \Cref{table:gsr_results}. For $x_d^{\rm fid}=100$, the signal is more coherent and localised within the folded and intermediate triangle regions, matching the tighter uncertainties obtained for $C$ and $\eta_f$.

The comparison illustrates the complementary scale coverage of single-dish and interferometric surveys. SKAO samples large-scale modes directly, but foreground removal and the loss of radial modes suppress regions of triangle space relevant to the feature signal. Conversely, HIRAX features a more restricted window at low redshifts but probes a larger number of small- and intermediate-scale transverse modes. When this accessible domain overlaps with high-SNR oscillatory bands, the resulting bispectrum uncertainties tighten, accounting for the lower forecasted uncertainties from HIRAX relative to SKAO despite the removal of portions of its triangle domain.

The dependence of feature uncertainties on frequency and phase is determined by the alignment of the oscillatory bands within triangle space relative to the survey window. Modifications to these parameters shift the bands; if the high-SNR regions coincide with scales excluded by foreground cuts, beam effects, or the interferometric wedge, the uncertainties degrade. Conversely, alignment with well-measured configurations increases the relative sensitivity contribution from the bispectrum. The survey window thus acts as a filter that determines the sensitivity to specific feature frequencies and phases.

Folded and elongated configurations contribute significantly across all templates, particularly within the sharp-feature model, though the signal extends beyond a single geometric limit. Intermediate and equilateral-like configurations also contain non-negligible information, indicating that detection strategies limited to a single triangle shape omit parts of the oscillatory signal. Maintaining broad triangle-shape coverage adapted to the specific survey window ensures optimal information extraction: HIRAX probes small- and intermediate-scale configurations, whereas SKAO provides complementary sensitivity on larger scales. A joint analysis combining multiple redshift bins, survey windows, and an extended set of triangle shapes maximises sensitivity relative to an analysis optimised for a single configuration.

The frequency dependence is illustrated for the linear template in \Cref{fig:SN_lin_freq_SKAO,fig:SN_lin_freq_HIRAX}. An increase in $\omega_{\rm lin}$ reduces the spacing between the oscillatory bands and shifts them across the triangle domain. This alters the overlap with the survey selection functions: in SKAO, the high-SNR regions track primarily across folded and elongated configurations, whereas in HIRAX this translation is modulated by the interferometric window and wedge. This mechanism drives the frequency dependence of the amplitude uncertainties and the survey-specific nature of the optimal triangle configurations.

\subsection{Impact of non-Gaussian covariances on feature detection} \label{sec:full_bispectrum_covariance}
As discussed in~\Cref{sec:method_HI}, our baseline bispectrum forecasts adopt the Gaussian covariance and neglect non-Gaussian contributions. To assess the impact of the latter, we repeat the forecasts including an approximate non-Gaussian covariance. For this test we fix $k_{\rm max}=0.2\,h\,{\rm Mpc}^{-1}$, which keeps the calculation computationally tractable while still probing the relevant regime for the bispectrum analysis. For the non-Gaussian part, we retain the dominant bispectrum--bispectrum and power-spectrum--trispectrum contributions, following Refs.~\cite{Barreira:2019icq,Biagetti:2021tua,Salvalaggio:2024vmx},
\begin{align}
    C^{\rm B}_{\rm NG}(\bk_i,\bk_j)
    =
    2\,\frac{(2\pi)^3}{V_s\,V_{123}^i\,V_{123}^j}
    \Big[
    &\delta_{k_1^i k_1^{j}}\,
    U(k_1^i,k_1^j)\,
    B_{\rm HI}(\bk_1^{j},\bk_2^i,\bk_3^i)\,
    B_{\rm HI}(\bk_1^i,\bk_2^{j},\bk_3^{j})
    \notag\\
    &+8\,{\rm perm.}
    \Big],
    \label{eq:Cov_NG}
\end{align}
with
\begin{equation}
    U(k_1^i,k_1^j)
    =
    16\pi^3\,
    k_2^i k_3^i k_2^j k_3^j
    (\Delta k)^5 .
\end{equation}
For flattened and open triangle configurations we apply the analytic corrections of Ref.~\cite{Biagetti:2021tua}. Since the instrumental noise of the HI instrument is taken to be Gaussian~\cite{Bull2015}, it enters only the Gaussian covariance through the observed power spectrum $P_{\rm HI}$ and does not contribute to $C^{\rm B}_{\rm NG}$. We emphasise that this prescription is intended as a diagnostic of the impact of non-Gaussian covariance terms, not as an exact evaluation of the full bispectrum covariance.

\begin{figure}[t]
    \centering
    \includegraphics[width=0.49\textwidth]{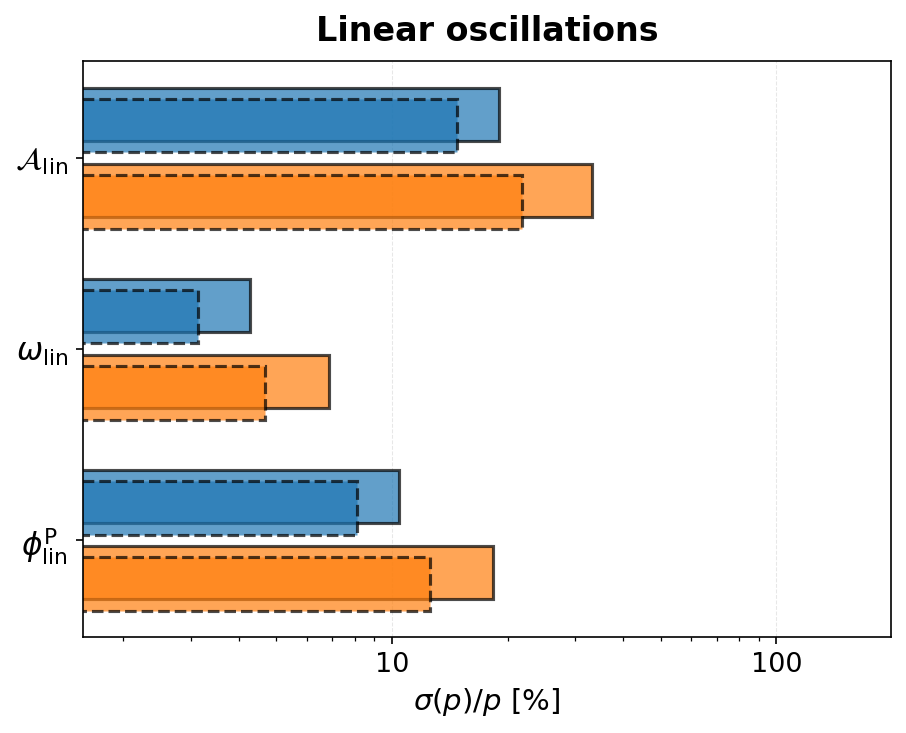}
    \hfill
    \includegraphics[width=0.49\textwidth]{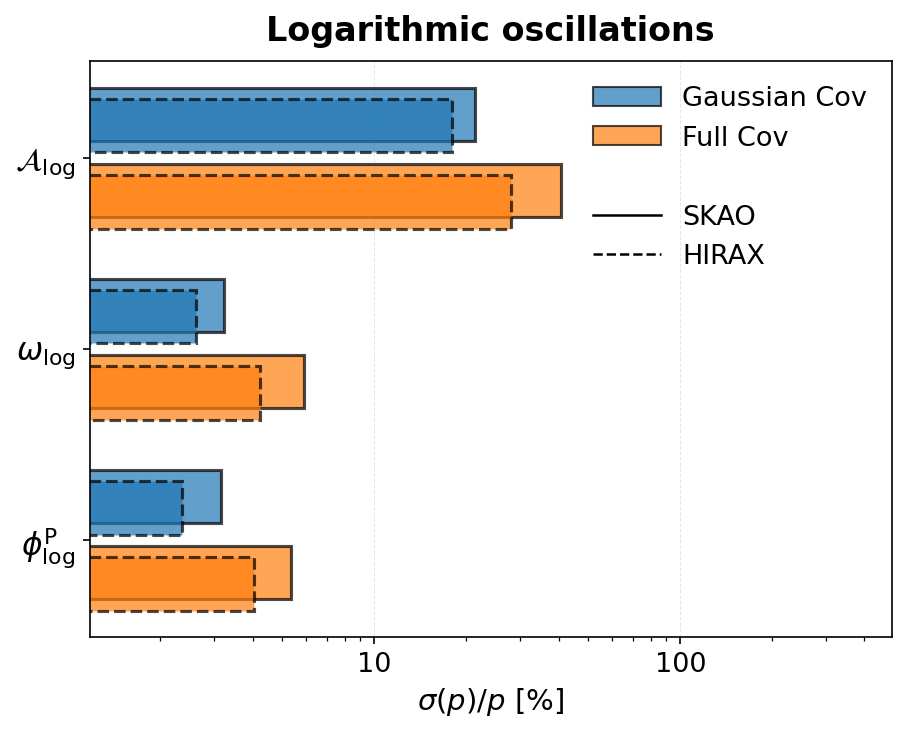}

    \vspace{0.3cm}

    \includegraphics[width=0.49\textwidth]{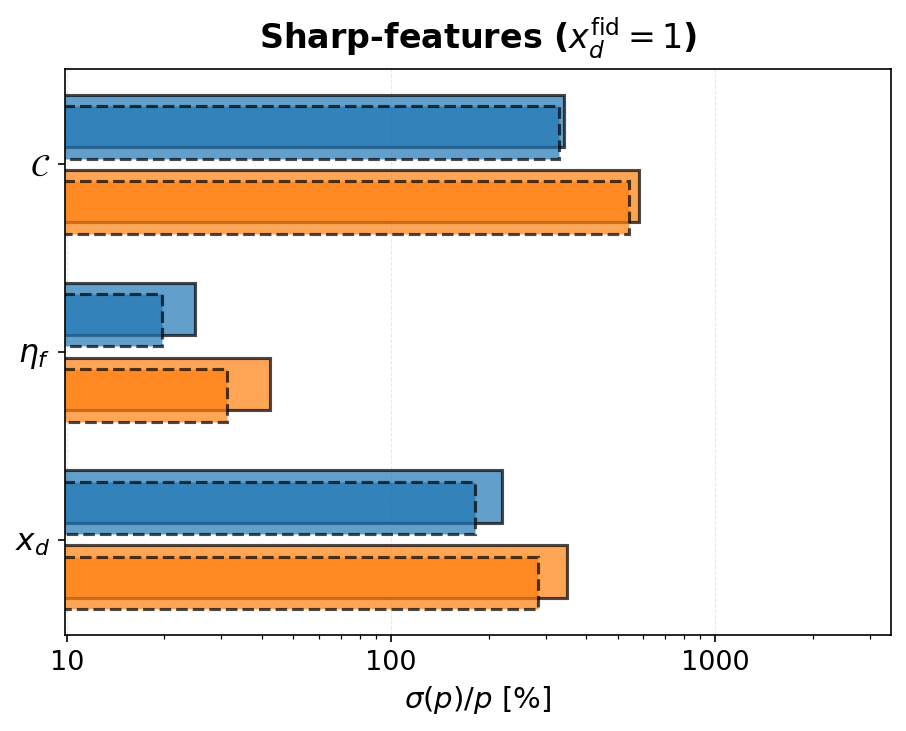}
    \hfill
    \includegraphics[width=0.49\textwidth]{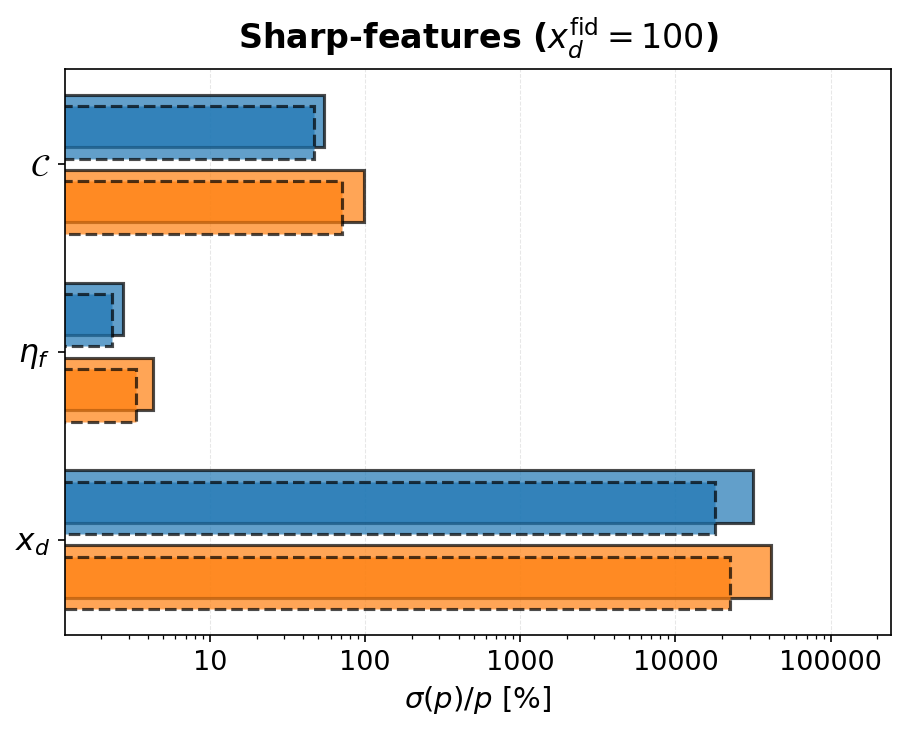}

    \caption{Bispectrum-only constraints for a fixed $k_{\rm max}=0.2\,h\,{\rm Mpc}^{-1}$, comparing results obtained with the Gaussian covariance and with the full covariance. The panels show the linear oscillatory template, logarithmic oscillatory template, and the sharp-features model with $x_d^{\rm fid}=1$ and $x_d^{\rm fid}=100$, respectively.}
    \label{fig:BS_covariance_comp}
\end{figure}

Including the non-Gaussian covariance systematically weakens the bispectrum constraints, as expected. For the linear template, the constraints on $\mathcal{A}_{\rm lin}$, $\omega_{\rm lin}$ and $\phi_{\rm lin}^{\rm P}$ degrade by $60$--$75\%$ for SKAO and $50$--$55\%$ for HIRAX. The logarithmic template shows somewhat larger degradations: $70$--$90\%$ for SKAO and $55$--$75\%$ for HIRAX. The forecasts for the sharp-feature model follow the same pattern. For $x_d^{\rm fid}=1$, the constraints on $\mathcal{C}$, $\eta_f$ and $x_d$ weaken by $60$--$70\%$ for SKAO and $55$--$65\%$ for HIRAX. For $x_d^{\rm fid}=100$ the response becomes more parameter-dependent: the amplitude degrades by $\sim 80\%$ (SKAO) and $\sim 50\%$ (HIRAX), $\eta_f$ by $\sim 55\%$ and $\sim 45\%$, while the already weak constraint on $x_d$ moves by only $\sim 30\%$ and $\sim 25\%$.

This test confirms that non-Gaussian covariance terms are not negligible for bispectrum forecasts on the feature parameters considered here. The qualitative picture nonetheless survives. HIRAX remains more constraining than SKAO for the feature models considered here, and the bispectrum retains useful sensitivity, particularly to the oscillation amplitude and frequency.

\section{Conclusion} \label{sec:conclusion}

We have forecasted the power spectrum and bispectrum potential of future
HI intensity mapping (SKAO in single-dish mode, HIRAX as an interferometer) and CMB
(SO+LiteBIRD) measurements to constrain primordial oscillatory features, for linear and
logarithmic templates, as well as a sharp-feature model, under optimistic and pessimistic foreground assumptions.

HIRAX consistently outperforms SKAO: in the optimistic scenario, $P+B$ amplitude
constraints are $60$--$70\%$ tighter for the linear and logarithmic templates, and
$40$--$50\%$ tighter for the sharp feature ($x_d^{\rm fid}=1$).  This follows from the
interferometer's access to the transverse intermediate and small scales where
oscillatory signals are most distinguishable from a smooth spectrum, while SKAO's
single-dish mode is more exposed to the loss of radial modes under foreground cuts.
The bispectrum is not a marginal addition: combining $P+B$ tightens the amplitude
constraints by $\sim30\%$ (SKAO) and $35$--$40\%$ (HIRAX) relative to the power
spectrum alone, with comparable gains for the phase, and in several cases the
bispectrum alone carries more information on the power-spectrum feature parameters
than the power spectrum itself.

A central result of this work is that this gain is not driven solely by the
primordial bispectrum contribution: the primordial bispectrum amplitude itself
remains essentially undetected. Instead, most of the improvement on the
power-spectrum feature parameters comes from the late-time gravitational term in the
tree-level bispectrum, usually treated as a nuisance in primordial Universe studies. Because this term inherits the oscillatory structure of the feature power
spectrum (\cref{eq:Bgs}), it remains sensitive to the primordial signal even when the
primordial bispectrum itself is undetected. The bispectrum is therefore an
independent and highly informative probe on the power-spectrum feature parameters in
its own right, justifying its inclusion in feature searches generally.

The CMB is competitive for the frequency parameters $\omega_X$ and $\eta_f$, but is
generally outperformed by HIRAX $P+B$ on the amplitude, by roughly $75\%$ for the
linear template, and provides no useful constraint on the oscillation phase, where
the three-dimensional LSS data dominate through their many Fourier modes and triangle
configurations.  The joint HI+CMB combination is most useful for SKAO under
pessimistic foregrounds, while for HIRAX the gain is smaller since the HI constraints
already dominate.  

Constraining power is also frequency-dependent across all observables: amplitude
constraints plateau for $\omega_{\rm lin}\gtrsim10$ and show a localised degradation
near $\omega_{\rm lin}\sim7$--$9$ where the feature overlaps the BAO scale; the
sharp-feature model shows the analogous pattern, with $x_d^{\rm fid}=1$ degrading at
large $\eta_f$ and $x_d^{\rm fid}=100$ instead reaching a broad plateau. For the bispectrum, this behaviour is further shaped by how the signal-to-noise is distributed across triangle configurations (\Cref{fig:SN_lin,fig:SN_log,fig:SN_gsr_xd_1,fig:SN_gsr_xd_100} and \cref{app:sn_triangles}). Folded and elongated configurations carry much of the signal, but no single shape dominates. Detection strategies restricted to a narrow class of triangles are therefore sub-optimal, while the survey window (including foreground and
wedge cuts) effectively filters which oscillation frequencies and phases are most accessible.

Including the non-Gaussian bispectrum covariance (at fixed $k_{\rm max}=0.2\,h\,{\rm Mpc}^{-1}$) degrades the bispectrum constraints on the power-spectrum feature parameters by $55$--$85\%$ for both surveys, while $f_{\rm NL}^X$ and $\phi_X^{\rm B}$ — constrained through the primordial part of the bispectrum — shift by only a few percent. The qualitative picture is unchanged: HIRAX remains more constraining than SKAO, and $P+B$ retains significant sensitivity to the oscillatory signal.

Overall, joint HI IM and CMB analyses that exploit both the power spectrum and the bispectrum provide a powerful and complementary strategy for detecting primordial features. Upcoming interferometric surveys such as HIRAX can reach few-percent precision on feature amplitudes and frequencies across a wide range of oscillation scales, surpassing the CMB in several cases. When combined with CMB information, they yield the tightest constraints among the probes considered here. The bispectrum plays a central role, both through its primordial component and, distinctively for feature models, through its late-time gravitational contribution. A more complete treatment of bispectrum covariance, foreground residuals, and instrumental systematics will be essential for fully realising this potential.

\acknowledgments
The authors thank Daan Meerburg for useful comments on the manuscript. DK acknowledges support by the MUR PRIN2022 Project “BROWSEPOL: Beyond standaRd mOdel With coSmic microwavE background POLarization”-2022EJNZ53 financed by the European Union- Next Generation EU. MB acknowledges financial support from the COSMOS network through the ASI (Italian Space Agency) Grants 2016-24-H.0, 201624-H.1-2018, 2020-9-HH.0 (participation in LiteBIRD phase A).

\bibliographystyle{JHEP}
\bibliography{references}

\appendix
\section{Features with larger amplitude}
\label{app:Amplx10}

For completeness, Tables~\ref{tab:lin_log_results_Amplx10} and \ref{tab:gsr_results_Amplx10} report the corresponding forecasts obtained when the primordial feature amplitude is increased by a factor of ten relative to the fiducial setup. The former shows the linear and logarithmic templates, while the latter presents the sharp-feature results for $x_d=1$ and $x_d=100$.

\begin{table*}[h!]
\centering
\resizebox{\textwidth}{!}{
\begin{tabular}{l ccc ccc c}
\hline\hline
 & \multicolumn{3}{c}{\bf SKAO} 
 & \multicolumn{3}{c}{\bf HIRAX}
 & {\bf CMB} \\
Parameter & $P$ & $B$ & $P+B$ & $P$ & $B$ & $P+B$ & $P$ \\
\hline
\\[-0.8em]
\multicolumn{8}{c}{\bf Linear, ${\cal A}_X=0.1$} \\[0.3em]
\hline
$\sigma({\cal A}_{\rm lin})/{\cal A}_{\rm lin}$ 
& 0.02 (0.05) & 0.02 (0.12) & 0.01 (0.04) 
& 0.009 (0.01) & 0.007 (0.01) & 0.005 (0.008)
& 0.02 \\

$\sigma(\omega_{\rm lin})/\omega_{\rm lin}$ 
& 0.008 (0.01) & 0.008 (0.03) & 0.006 (0.01) 
& 0.004 (0.006) & 0.003 (0.007) & 0.002 (0.004)
& 0.001 \\

$\sigma(\phi_{\rm lin}^{\rm P})$ 
& 0.01 (0.02) & 0.01 (0.05) & 0.007 (0.02) 
& 0.004 (0.005) & 0.003 (0.006) & 0.002 (0.004)
& -- \\

$\sigma(f_{\rm NL}^{\rm lin})/f_{\rm NL}^{\rm lin}$ 
& -- & 1.7 (6.7) & 1.7 (6.7) 
& -- & 0.72 (1.45) & 0.72 (1.45)
& -- \\

$\sigma(\phi_{\rm lin}^{\rm B})$ 
& -- & 0.35 (1.37) & 0.35 (1.35) 
& -- & 0.15 (0.29) & 0.15 (0.29)
& -- \\[0.3em]
\hline
\\[-0.8em]
\multicolumn{8}{c}{\bf Logarithmic, ${\cal A}_X=0.1$} \\[0.3em]
\hline
$\sigma({\cal A}_{\rm log})/{\cal A}_{\rm log}$ 
& 0.02 (0.05) & 0.02 (0.11) & 0.02 (0.04)
& 0.009 (0.01) & 0.008 (0.02) & 0.005 (0.008)
& 0.02 \\

$\sigma(\omega_{\rm log})/\omega_{\rm log}$ 
& 0.003 (0.009) & 0.003 (0.02) & 0.002 (0.007)
& 0.002 (0.004) & 0.001 (0.004) & 0.001 (0.003)
& 0.002 \\

$\sigma(\phi_{\rm log}^{\rm P})$ 
& 0.01 (0.03) & 0.01 (0.06) & 0.010 (0.02)
& 0.008 (0.010) & 0.005 (0.01) & 0.004 (0.007)
& -- \\

$\sigma(f_{\rm NL}^{\rm log})/f_{\rm NL}^{\rm log}$ 
& -- & 1.9 (12.2) & 1.8 (10.4)
& -- & 0.81 (1.88) & 0.80 (1.81)
& -- \\

$\sigma(\phi_{\rm log}^{\rm B})$ 
& -- & 0.38 (2.52) & 0.37 (1.88)
& -- & 0.16 (0.38) & 0.16 (0.37)
& -- \\[0.3em]
\hline\hline
\end{tabular}
}
\caption{Same as Table~\ref{table:lin_log_results}, but for a primordial feature amplitude ten times larger than in the fiducial linear and logarithmic templates. Values in parentheses correspond to the pessimistic foreground cut. The CMB column lists angular power spectrum uncertainties from SO+LiteBIRD.}
\label{tab:lin_log_results_Amplx10}
\end{table*}

\begin{table*}[h!]
\centering
\resizebox{\textwidth}{!}{
\begin{tabular}{l ccc ccc c}
\hline\hline
 & \multicolumn{3}{c}{\bf SKAO}
 & \multicolumn{3}{c}{\bf HIRAX}
 & {\bf CMB} \\
Parameter & $P$ & $B$ & $P+B$ & $P$ & $B$ & $P+B$ & $P$ \\
\hline
\\[-0.8em]
\multicolumn{8}{c}{$\mathbf{Sharp\ feature}\,(x_d=1)$, ${\cal C}=0.1$} \\[0.3em]
\hline
$\sigma({\cal C})/{\cal C}$
& 0.36 (0.37) & 0.37 (0.39) & 0.25 (0.26)
& 0.30 (0.30) & 0.17 (0.18) & 0.14 (0.15)
& 0.37 \\

$\sigma(\eta_f)/\eta_f$
& 0.027 (0.028) & 0.027 (0.029) & 0.018 (0.019)
& 0.024 (0.024) & 0.011 (0.012) & 0.010 (0.010)
& 0.030 \\

$\sigma(x_d)/x_d$
& 0.22 (0.23) & 0.24 (0.25) & 0.16 (0.16)
& 0.15 (0.15) & 0.094 (0.097) & 0.078 (0.081)
& 0.22 \\[0.3em]

\hline
\\[-0.8em]
\multicolumn{8}{c}{$\mathbf{Sharp\ feature}\,(x_d=100)$, ${\cal C}=0.1$} \\[0.3em]
\hline
$\sigma({\cal C})/{\cal C}$
& 0.055 (0.059) & 0.054 (0.061) & 0.038 (0.041)
& 0.033 (0.033) & 0.019 (0.020) & 0.016 (0.016)
& 0.029 \\

$\sigma(\eta_f)/\eta_f$
& 0.004 (0.004) & 0.004 (0.005) & 0.003 (0.003)
& 0.002 (0.002) & 0.002 (0.002) & 0.001 (0.001)
& 0.001 \\

$\sigma(x_d)/x_d$
& 23.14 (23.57) & 33.45 (35.05) & 18.37 (18.86)
& 3.99 (4.01) & 4.03 (4.07) & 2.55 (2.56)
& 8.61 \\[0.3em]

\hline\hline
\end{tabular}
}
\caption{Same as Table~\ref{table:gsr_results}, but for a primordial feature amplitude ten times larger than in the fiducial sharp-feature templates with $x_d^{\rm fid}=1$ and $x_d^{\rm fid}=100$. Values in parentheses correspond to the pessimistic foreground cut. The CMB column lists angular power spectrum uncertainties from SO+LiteBIRD.}
\label{tab:gsr_results_Amplx10}
\end{table*}

\section{Frequency dependence of bispectrum triangle sensitivity} \label{app:sn_triangles}
In this appendix we show the angularly averaged bispectrum SNR contribution across triangle configurations. These plots are intended as a diagnostic of how the oscillatory feature signal is distributed over triangle space and how this distribution changes with frequency, survey geometry, and foreground cuts.

\begin{figure}[h!]
    \centering
    {\large\bfseries\boldmath $\omega_{\rm lin}=10$}\par
    \includegraphics[width=\textwidth]{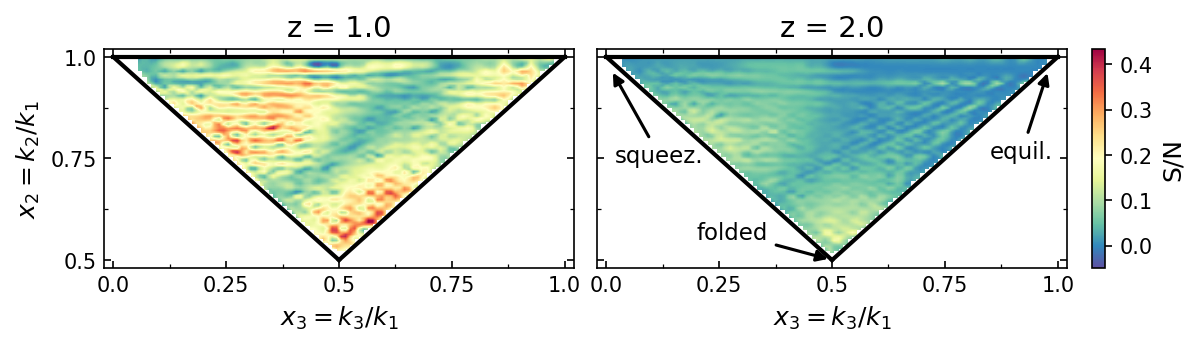}
    {\large\bfseries\boldmath $\omega_{\rm lin}=20$}\par
    \includegraphics[width=\textwidth]{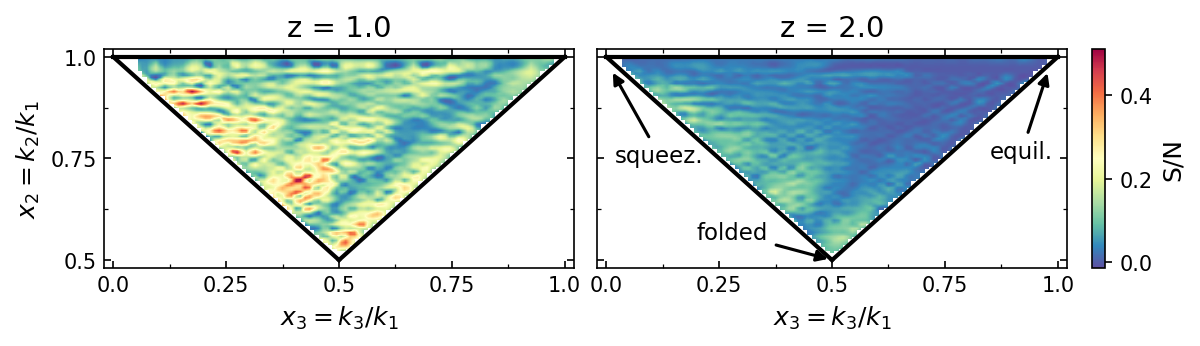}
    {\large\bfseries\boldmath $\omega_{\rm lin}=40$}\par
    \includegraphics[width=\textwidth]{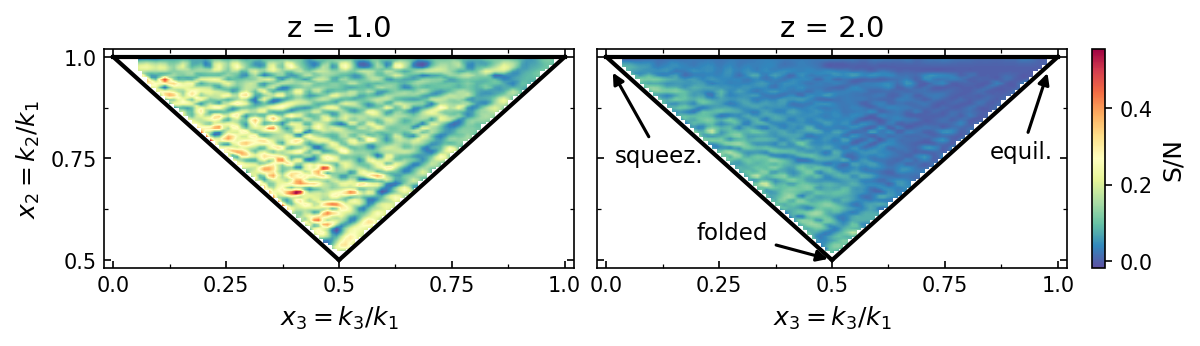}
    \caption{Angular-averaged bispectrum SNR contribution as a function of triangle shape for the linear feature model in SKAO. From top to bottom, the panels show increasing oscillation frequency, $\omega_{\rm lin}=10,20,40$. Increasing the frequency shifts the oscillatory high SNR bands across triangle space, changing their overlap with the survey window and foreground cuts.}
    \label{fig:SN_lin_freq_SKAO}
\end{figure}

\begin{figure}[h!]
    \centering
    {\large\bfseries\boldmath $\omega_{\rm lin}=10$}\par
    \includegraphics[width=\textwidth]{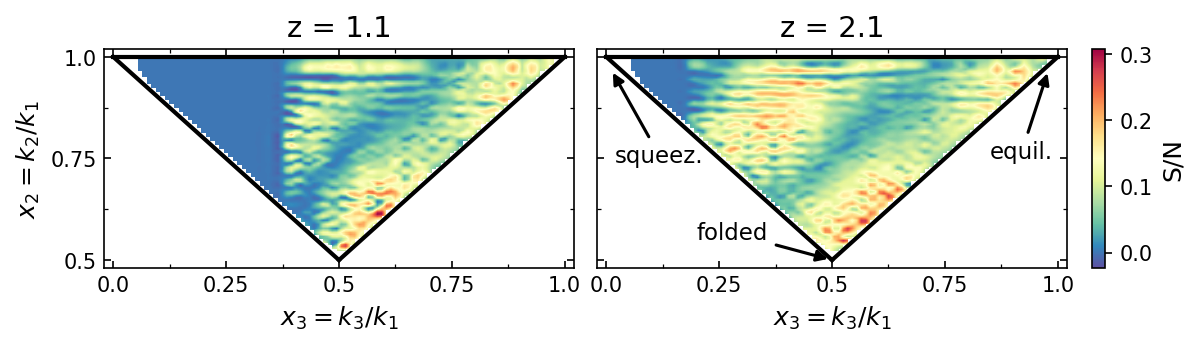}
    {\large\bfseries\boldmath $\omega_{\rm lin}=20$}\par
    \includegraphics[width=\textwidth]{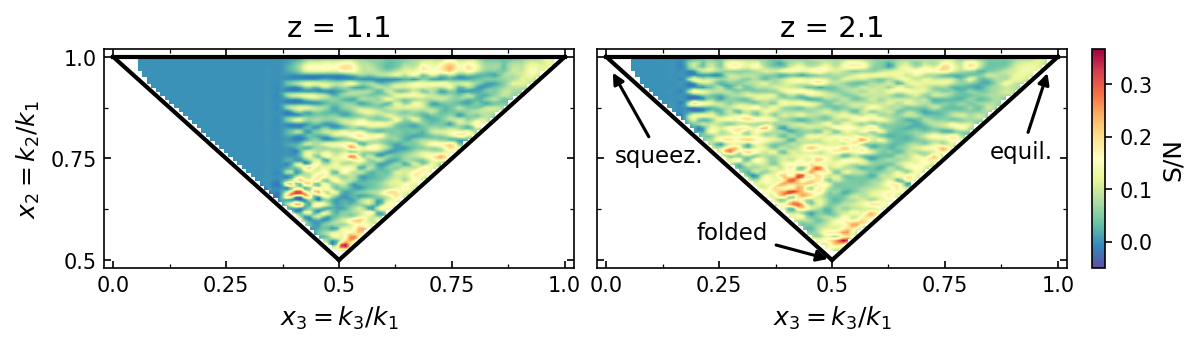}
    {\large\bfseries\boldmath $\omega_{\rm lin}=40$}\par
    \includegraphics[width=\textwidth]{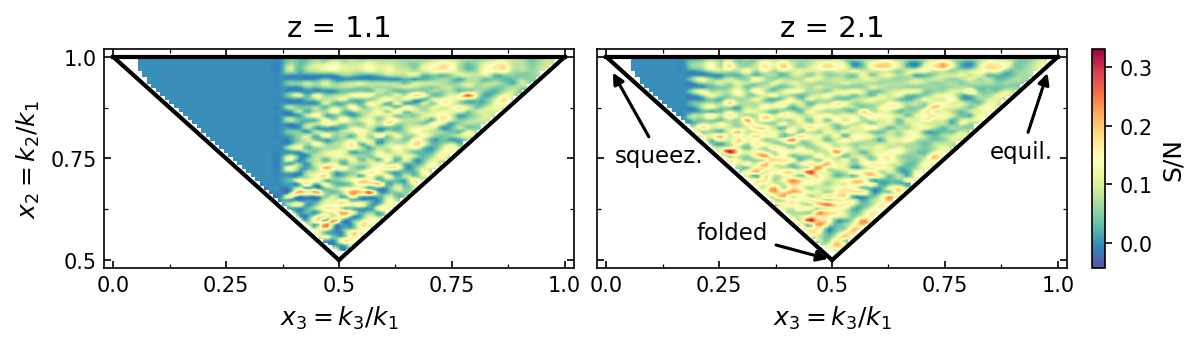}
    \caption{As in \Cref{fig:SN_lin_freq_SKAO}, but for HIRAX. The blank region corresponds to triangle configurations removed by the interferometric survey window and foreground wedge. As $\omega_{\rm lin}$ increases, the oscillatory bands become more closely spaced, so the measured signal depends sensitively on whether the high SNR regions fall within the accessible part of triangle space.}
    \label{fig:SN_lin_freq_HIRAX}
\end{figure}

\end{document}